%% file: main.tex
\documentclass[fleqn,10pt]{wlpeerj}

\usepackage{algorithm}
\usepackage{algpseudocode}
\usepackage{booktabs}
\usepackage{multirow}
\usepackage{float}
\usepackage{microtype}
\usepackage{hyperref}

\DeclareUnicodeCharacter{0394}{$\Delta$}

\title{Markovian Promoter Models: A Mechanistic Alternative to Hill Functions in Gene Regulatory Networks}

\author[1]{Tianyu Wu}

\affil[1]{Center for Biophysics and Quantitative Biology, University of Illinois at Urbana–Champaign, Urbana, IL, USA}
\affil[2]{National Science Foundation Science and Technology Center for Quantitative Cell Biology, Beckman Institute for Advanced Science and Technology, University of Illinois at Urbana–Champaign, Urbana, IL, USA}

\corrauthor[1]{Tianyu Wu}{tianyu16@illinois.edu}

\begin{document}

\flushbottom
\thispagestyle{empty}

\begin{abstract}
\input{abstract}
\end{abstract}
\maketitle

\section*{Introduction}
\input{introduction}

\section*{Methods}
\input{methods}

\section*{Results}
\input{results}

\section*{Discussion}

\input{discussion}
\input{discussion_wholecell}

\section*{Conclusion}
\input{conclusion}

\section*{Code Availability}
Code is available in GitHub Repository: https://github.com/forxhunter/MM\_Promoter .

\bibliography{references}

\end{document}

%% file: abstract.tex

Gene regulatory networks are typically modeled using ordinary differential equations (ODEs) with phenomenological Hill functions to represent transcriptional regulation. While computationally efficient, Hill functions lack mechanistic grounding and cannot capture stochastic promoter dynamics. We present a hybrid Markovian-ODE framework that explicitly models discrete promoter states while maintaining computational tractability. Uniquely, we parameterize this model using fractional dwell times derived from ChEC-seq data, enabling the inference of in vivo kinetic rates from steady-state chromatin profiling. Our approach tracks individual transcription factor binding events as a continuous-time Markov chain, linked to deterministic molecular dynamics. We validate this framework on seven gene regulatory systems spanning basic to advanced complexity: the GAL system, repressilator, Goodwin oscillator, toggle switch, incoherent feed-forward loop, p53-Mdm2 oscillator, and NF-$\kappa$B pathway. Comparison with stochastic simulation algorithm (SSA) ground truth demonstrates that Markovian promoter models achieve similar accuracy to full stochastic simulations while being 10-100$\times$ faster. Our framework provides a mechanistic foundation for gene regulation modeling and enables investigation of promoter-level stochasticity in complex regulatory networks.

%% file: introduction.tex
Gene regulatory networks (GRNs) govern fundamental cellular processes including development, homeostasis, and response to environmental stimuli \citep{alon2007network, davidson2010emerging}. As whole-cell modeling efforts advance \citep{karr2012whole, macklin2014simultaneous}, there is increasing need to integrate gene regulation with stochastic chemical kinetics frameworks such as the Chemical Master Equation (CME) \citep{gillespie1992rigorous, van2007stochastic}. However, this integration faces a critical challenge: gene regulation is inherently nonlinear and involves complex binding kinetics that are difficult to parameterize from available experimental data \citep{bintu2005transcriptional,wuSpatialHeterogeneityAlters2025}.

The dominant modeling paradigm employs ordinary differential equations (ODEs) with Hill functions to represent transcriptional regulation \citep{HillThePE, alon2007network}. In this framework, the rate of protein production is described by $d[P]/dt = k_{trans} \cdot [TF]^n/(K_d^n + [TF]^n) - k_{deg}[P]$, where $[P]$ is protein concentration, $[TF]$ is transcription factor concentration, $k_{trans}$ is the maximum transcription rate, $K_d$ is the dissociation constant, and $n$ is the Hill coefficient. While computationally efficient and widely used in systems biology \citep{elowitz2000synthetic, gardner2000construction, tyson2003sniffers}, Hill functions present difficulties for stochastic simulation like whole-cell modeling. First, they are deterministic and cannot capture promoter-level noise, which is increasingly recognized as functionally important for cellular decision-making \citep{elowitz2002stochastic, raj2008nature, sanchez2013regulation}. Second, the Hill coefficient $n$ does not correspond to a true reaction order in mass-action kinetics, making it incompatible with CME frameworks that require polynomial propensities \citep{gillespie1977exact, anderson2007error}. Third, detailed binding/unbinding rate constants ($k_{on}$, $k_{off}$) for each transcription factor-DNA interaction are rarely available \citep{rohs2010origins}, yet proteomic and transcriptomic data are increasingly abundant \citep{vogel2012insights, li2014quantifying}. Finally, Hill functions lack mechanistic grounding, obscuring the relationship between molecular binding events and system-level dynamics \citep{bintu2005transcriptional, buchler2003molecular}. The Michaelis-Menten formalism, originally developed for enzyme kinetics \citep{michaelis1913kinetics}, has been adapted for gene regulation with the form $d[mRNA]/dt = V_{max} \cdot [TF]/(K_M + [TF]) - k_{deg}[mRNA]$. This approach suffers from similar limitations as Hill functions: it is deterministic and cannot capture stochastic promoter switching \citep{peccoud1995markovian}, it lacks cooperativity and requires ad hoc modifications to capture ultrasensitive responses \citep{goldbeter1996biochemical}, and the fractional form violates mass-action kinetics \citep{frankInputoutputRelationsBiological2013}. Moreover, the steady-state assumption underlying Michaelis-Menten kinetics may not hold for gene regulation, where binding/unbinding can occur on similar timescales as transcription \citep{larson2013real, fukaya2016enhancer}. Recent approaches have proposed using neural networks to learn regulatory functions from data \citep{chen2018neural, rackauckas2020universal}, where the production rate is given by $d[P]/dt = NN([TF_1], [TF_2], \ldots; \theta) - k_{deg}[P]$ with $NN(\cdot; \theta)$ representing a neural network with parameters $\theta$. While flexible, this approach has critical drawbacks for whole-cell modeling. Neural networks are black boxes that provide no mechanistic insight into regulatory logic \citep{lipton2018mythos}, they require extensive training data and have many parameters with no biological meaning \citep{samekExplainableArtificialIntelligence2017,raissiPhysicsinformedNeuralNetworks2019}, they are deterministic and thus cannot capture stochasticity, their outputs are not polynomial in molecule numbers making them CME-incompatible, they may fail to generalize beyond training data particularly for perturbations or mutations \citep{xu2020neural}, and evaluating neural networks at each simulation step is computationally expensive for large-scale models, which scales poorly for large biochemical systems involving thousands of reactions and frequent rate evaluations\citep{chen2018neural,gholamiANODEUnconditionallyAccurate2019,massaroliDissectingNeuralODEs2020}.

Some approaches attempt to approximate gene regulation using simple mass-action reactions \citep{cao2005efficient}, explicitly modeling binding ($TF + DNA \xrightarrow{k_{on}} TF:DNA$), unbinding ($TF:DNA \xrightarrow{k_{off}} TF + DNA$), and transcription ($TF:DNA \xrightarrow{k_{trans}} TF:DNA + mRNA$). While CME-compatible, this approach has severe limitations. Ignoring cooperative binding, multiple binding sites, and combinatorial logic loses essential regulatory features \citep{buchler2003molecular, bintu2005transcriptional}. Simplifications at the promoter level propagate through the network, leading to qualitatively incorrect behavior \citep{thomas2013phenotypic}. Simple binding cannot generate the sharp dose-response curves observed experimentally \citep{ferrell1996tripping}, and explicitly modeling all TF-DNA complexes for multiple genes becomes computationally intractable \citep{drawert2012stochastic}. In summary, the fundamental problem with existing approaches is their lack of generalizability. Each regulatory system requires custom modifications such as different Hill coefficients or neural network architectures. Different systems use different modeling paradigms, hindering integration into whole-cell models. Parameters fitted to one condition often fail to predict behavior under perturbations \citep{gutenkunst2007universally}.

The Gillespie Stochastic Simulation Algorithm (SSA) \citep{gillespie1977exact} and the Chemical Master Equation \citep{gillespie1992rigorous} provide rigorous frameworks for stochastic chemical kinetics. By explicitly tracking all molecular events, SSA captures intrinsic noise from discrete molecular events \citep{elowitz2002stochastic}, rare stochastic transitions in bistable systems \citep{gardner2000construction}, and promoter bursting and transcriptional noise \citep{raj2006stochastic}. However, explicitly modeling all binding/unbinding events for gene regulation becomes computationally prohibitive in whole-cell models. The computational cost scales as $O(N_{genes} \times N_{TF} \times \bar{n})$ where $N_{genes}$ is the number of genes, $N_{TF}$ is the number of transcription factors, and $\bar{n}$ is the average number of binding sites \citep{drawert2012stochastic}. Gene regulation involves fast binding events on the timescale of seconds and slow expression dynamics on the timescale of minutes to hours, requiring prohibitively small time steps \citep{haseltine2002approximate}. Furthermore, detailed $k_{on}$ and $k_{off}$ values for every TF-DNA interaction are rarely available \citep{rohs2010origins}.

We present a hybrid framework that bridges gene regulation and stochastic chemical kinetics by coarse-graining detailed binding kinetics into discrete promoter states while maintaining proper stochastic dynamics and reaction orders. The framework maintains several key principles. All reactions remain 0th, 1st, or 2nd order, compatible with standard CME frameworks \citep{gillespie1977exact, anderson2007error}. Promoter states are discrete and stochastic, capturing gene expression noise at its source \citep{peccoud1995markovian, sanchez2013regulation}. The model is grounded in binding kinetics \citep{ackers1982quantitative} but can be parameterized from accessible transcriptomic and proteomic data \citep{vogel2012insights} rather than requiring detailed binding constants. Gene regulatory modules can be seamlessly incorporated into existing whole-cell CME models without modifying the core simulation engine \citep{karr2012whole}. Cooperative binding \citep{ackers1982quantitative} and multi-input logic \citep{buchler2003molecular} emerge naturally from the Markovian state space without violating mass-action kinetics. Meanwhile, we also validate the methods by leverage high-throughput experimental ChEC-seq (Chromatin Endogenous Cleavage) data \citep{rossiHighresolutionProteinArchitecture2021a}, which offers higher spatial and temporal resolution than ChIP-seq. By interpreting ChEC-seq signal intensities as fractional dwell times, we can infer effective in vivo kinetic rates for promoter states \citep{vanbelzenChromatinEndogenousCleavage2024a}, turning static binding maps into dynamic kinetic models.

The key insight is to coarse-grain the detailed binding kinetics into a discrete promoter state $S \in \{0, 1, \ldots, n\}$ representing the number of bound transcription factors. This allows gene regulation to be treated as a combination of promoter transitions (1st and 2nd order reactions between promoter states), transcription (state-dependent 0th order production), and post-transcriptional processes (standard 1st and 2nd order reactions). All reactions maintain proper mass-action kinetics, enabling seamless integration with CME frameworks while capturing the essential stochastic and nonlinear features of gene regulation.

As shown in Table \ref{tab:comparison}, our approach is the only one that satisfies all criteria: stochastic, CME-compatible, mechanistic, scalable, and parameterizable from accessible data.

\begin{table}[ht]
\centering
\caption{Comparison of Gene Regulation Modeling Approaches}
\label{tab:comparison}
\begin{tabular}{p{2.5cm}p{1.8cm}p{1.8cm}p{1.8cm}p{2cm}p{2cm}}
\toprule
\textbf{Approach} & \textbf{Stochastic} & \textbf{CME Compatible} & \textbf{Mechanistic} & \textbf{Scalable} & \textbf{Parameterization} \\
\midrule
Hill functions & No & No & No & Yes & Transcriptomics \\
Michaelis-Menten & No & No & Partial & Yes & Transcriptomics \\
Neural Network ODE & No & No & No & Moderate & Extensive training \\
Simple mass-action & Yes & Yes & Partial & No & Binding assays \\
Full SSA & Yes & Yes & Yes & No & Binding assays \\
\textbf{Markovian promoters} & \textbf{Yes} & \textbf{Yes} & \textbf{Yes} & \textbf{Yes} & \textbf{Transcriptomics} \\
\bottomrule
\end{tabular}
\end{table}

This framework addresses a critical bottleneck in whole-cell modeling: how to incorporate complex gene regulation without losing stochastic effects that are functionally important for cellular decision-making \citep{elowitz2002stochastic, losick2008stochasticity}, without requiring extensive parameterization of binding kinetics that are experimentally inaccessible \citep{rohs2010origins}, without breaking compatibility with existing CME and SSA simulation frameworks \citep{drawert2012stochastic, sanft2011stochkit}, and without sacrificing computational efficiency needed for genome-scale models \citep{karr2012whole, macklin2014simultaneous}. We validate this approach on seven regulatory systems spanning basic to advanced complexity, demonstrating 10-100$\times$ speedup over full SSA while maintaining stochastic fidelity and proper reaction orders. Each system was selected to represent a distinct regulatory architecture and dynamic behavior, providing comprehensive validation of the framework's generalizability.

%% file: methods.tex
\subsection{Markovian Promoter Model}

\subsubsection{Single-Input Promoter}

Consider a gene promoter with $n$ binding sites for a transcription factor (TF). The promoter state is characterized by the number of bound TF molecules, $k \in \{0, 1, \ldots, n\}$. The state transitions follow a continuous-time Markov chain. Binding occurs with rate $k_{on} \cdot [TF] \cdot (n-k)$, while unbinding occurs with rate $k_{off} \cdot k$. Here, $k_{on}$ represents the binding rate constant and $k_{off}$ is the unbinding rate constant, with the dissociation constant given by $K_d = k_{off}/k_{on}$.

For cooperative binding, we introduce a cooperativity factor $q_r > 1$ that enhances the binding rate when sites are already occupied. Specifically, the binding rate becomes $k_{on} \cdot [TF] \cdot (n-k) \cdot q_r^{\mathbb{1}_{k>0}}$, reflecting the increased affinity of the specialized complex.

\subsubsection{Integration with Chemical Master Equation}

A key principle of our framework is that the Markovian promoter model maintains full compatibility with standard Chemical Master Equation (CME) and Stochastic Simulation Algorithm (SSA) frameworks by ensuring that all reactions preserve proper elementary orders.

In a CME framework, the system state $\mathbf{X}$ describes the copy numbers of all molecular species. The time evolution of the probability distribution $P(\mathbf{X}, t)$ is governed by propensities $a_{\mu}(\mathbf{X})$. Our approach augments the state space to include discrete promoter states $S_i \in \{0, 1, \ldots, n\}$ for each gene. Consequently, the augmented state vector includes both molecular counts and promoter indices.

The reaction set includes three distinct categories. First, promoter transitions involve state changes with no change in particle stoichiometry, such as binding events (second-order in TF concentration) and unbinding events (first-order in promoter state). Second, transcription is modeled as a state-dependent zero-order reaction, where the production rate $k_{trans}(S_i)$ depends explicitly on the discrete promoter state. Third, standard biochemical reactions such as translation, degradation, and complex formation proceed according to conventional first-order or second-order mass-action kinetics.

The critical advantage of this formulation is that all reactions maintain standard mass-action kinetics. The nonlinearity of gene regulation emerges naturally from the state-dependent transcription rate, rather than from non-physical fractional propensities often used in phenomenological models.

\subsubsection{Comparison with Hill Function Approach}

Hill function approaches in a CME context typically model transcription propensity as a fractional function of TF concentration, $k_{trans} \cdot X_{TF}^n / (K_d^n + X_{TF}^n)$. This formulation is problematic because the propensity is not a polynomial in molecule numbers, violating the fundamental assumption of mass-action kinetics. Furthermore, the effective reaction order becomes fractional and state-dependent, leading to a loss of mechanistic interpretation and incompatibility with standard SSA solvers. Our Markovian approach resolves these issues by decomposing the fractional response into discrete elementary steps, thereby preserving the stochastic nature of the system while recovering the dose-response behavior through emergent ensemble dynamics.

\subsubsection{Parameterization from Experimental Data}
A common challenge in mechanistic modeling is that detailed kinetic parameters for every TF-DNA interaction are rarely available. However, our framework allows parameters to be inferred from accessible experimental data types. Dose-response curves from transcriptomic data can be fitted to determine effective dissociation constants and cooperativity factors. Dynamic response data from time-series experiments provides timescale information to separate binding and unbinding rates. Furthermore, promoter switching rates can be constrained by noise characteristics derived from single-cell RNA-seq data, while structural data can inform the number of binding sites. This flexibility ensures that the model can be parameterized using the same data sources as Hill functions, but yields a mechanistically rigorous and CME-compatible representation.

\subsubsection{Promoter State Evolution}

The probability distribution over promoter states, denoted $\mathbf{p}(t)$, evolves according to the master equation $d\mathbf{p}/dt = \mathbf{Q}(t) \mathbf{p}$, where $\mathbf{Q}(t)$ is the time-dependent rate matrix. For small time steps $\Delta t$, the evolution can be computed using the matrix exponential. For computational efficiency when $\|\mathbf{Q}\| \Delta t \ll 1$, a first-order approximation is sufficient to update the probability distribution.

\subsubsection{Stochastic State Sampling}

At each time step, we sample a discrete promoter state from the evolved probability distribution using the inverse transform method. This step reintroduces the necessary stochasticity into the system, capturing the bursting behavior of gene expression.

\subsubsection{Transcription Rate}

The transcription rate is a function of the promoter occupancy. For an activator with $k$ bound sites, the rate is typically modeled as $k_{trans}(k) = k_{basal} + k_{max} \cdot f(k/n)$, where $f$ is an activity function. Common choices include proportional activity or a threshold function, depending on the specific regulatory logic of the gene.

\subsection{Multi-Input Promoter}

For promoters regulated by multiple TFs, the state space expands combinatorially. For $m$ TFs, the state is a vector characterizing the occupancy of each factor. We model the binding of each TF independently, with the total activity determined by a logic function $L(\mathbf{k})$. This function can encode complex regulatory logic, such as AND gates (requiring all factors to be bound), OR gates (requiring at least one factor), or custom logic like the specific repression conditions found in incoherent feed-forward loops.

\subsection{Hybrid ODE Coupling}

The Markovian promoter model is coupled with deterministic ODEs for molecular concentrations to form a hybrid simulation framework. In this scheme, the rate of mRNA production depends on the stochastically sampled promoter state $S(t)$, while translation and degradation processes follow coupled differential equations. This hybrid approach efficiently captures the slow, discrete dynamics of promoter switching and the fast, continuous dynamics of metabolic and signaling networks.

\subsection{Computational Implementation}

\subsubsection{Time Step Selection}
The communication time step $\Delta t$ is a critical parameter. It must be chosen to be sufficiently small to resolve the fastest promoter kinetics, typically satisfying $\Delta t \ll 1/k_{rate}$ for the fastest transition. We generally use $\Delta t$ values ranging from 0.001 to 0.01 minutes, depending on the specific system dynamics.

\subsubsection{Algorithm}

The simulation proceeds by initializing the probability distribution and molecular concentrations. At each time step, the rate matrix is updated based on current TF concentrations. The probability distribution is evolved, and a new discrete promoter state is sampled. This state determines the instantaneous transcription rate used to integrate the ODEs for mRNA and protein concentrations over the time interval.

\subsection{Validation Methodology}

We validate the Markovian promoter model by comparing it against two benchmarks: the exact Gillespie Stochastic Simulation Algorithm (SSA) as a ground truth for accuracy, and standard ODE models with Hill functions for behavioral reference. Validation metrics include trajectory correlation, promoter state agreement (quantified by Cohen's kappa), oscillation characteristics (period and amplitude), and computational execution time.

\input{methods_systems}

\input{methods_gal_conversion}

\input{methods_other_conversions}

\input{methods_dt_selection}
\input{methods_chec_markovian}

%% file: methods_systems.tex
\subsection{Model System Selection Criteria}

To comprehensively validate the Markovian promoter framework, we selected seven gene regulatory systems representing distinct network architectures, dynamic behaviors, and biological functions. The selection criteria were designed to ensure broad applicability and rigorous testing. We prioritized systems with diverse network topology, spanning simple negative feedback to complex multi-input networks \citep{alon2007network, milo2002network}. We included a range of dynamic behaviors such as bistability, oscillations, pulse generation, and homeostasis \citep{tyson2003sniffers}. Each system required well-characterized experimental data for parameter estimation and validation \citep{elowitz2000synthetic, gardner2000construction, lahav2004dynamics}. We ensured biological significance by selecting systems representing fundamental cellular processes like metabolism, cell cycle, stress response, and inflammation \citep{alon2007network}. Furthermore, we selected systems that are complex enough to challenge the framework but tractable enough for extensive simulation \citep{karr2012whole}, and which exhibit behaviors where stochasticity is functionally relevant \citep{elowitz2002stochastic, losick2008stochasticity}.

\subsubsection{Selected Systems and Rationale}

\paragraph{1. GAL System (Yeast Galactose Regulation)}

The GAL system serves as our primary validation case due to several key features. It exhibits multi-site binding, as GAL genes have 1-5 Gal4p binding sites, enabling validation of the framework's ability to handle variable binding site numbers \citep{lohr1995transcriptional}. It features dual regulation where Gal4p (activator) and Gal80p (repressor) provide a test of multi-input logic \citep{plattYeastGalactoseGenetic1998}. The system is extensively characterized with detailed kinetic and structural data available \citep{ramsey2006dual, venturelli2012synergistic}, and it represents a fundamental metabolic switch with biotechnological applications \citep{ostergaard2000metabolic}. Key features tested include variable binding site numbers (1-5 sites), cooperative binding, activator-repressor competition, and dose-response curves.

\paragraph{2. Repressilator (Synthetic Oscillator)}

The repressilator is a canonical synthetic biology circuit \citep{elowitz2000synthetic} chosen for its oscillatory dynamics, which tests the framework's ability to capture sustained oscillations. It consists of three genes in a cyclic repression topology, creating a ring of mutual repression. The system exhibits stochastic phase variations where oscillation period and amplitude vary stochastically \citep{elowitz2000synthetic}. Additionally, the inclusion of protein folding delay introduces an additional timescale, testing multi-timescale dynamics. This system allows us to test sustained oscillations, period and amplitude stochasticity, cyclic network topology, and protein maturation delays.

\paragraph{3. Goodwin Oscillator (Negative Feedback)}

The Goodwin oscillator is a classical model of circadian rhythms \citep{goodwin1965oscillatory} selected for its negative feedback oscillations, which are a fundamental regulatory motif in biology \citep{novak2008design}. It features nonlinear repression with a strong Hill coefficient, testing the cooperative binding implementation. The model is theoretically important and extensively studied in mathematical biology \citep{gonze2005goodwin}, representing core clock mechanisms \citep{goldbeter1996biochemical}. Key features tested include negative feedback loops, high cooperativity ($n \sim 10$), limit cycle oscillations, and robustness to parameter variations.

\paragraph{4. Toggle Switch (Bistable System)}

The toggle switch is another synthetic biology landmark \citep{gardner2000construction} chosen for its bistability, which tests rare stochastic switching between stable states. It features mutual repression with a symmetric network architecture. The system maintains state without continuous input, exhibiting memory \citep{ferrell2002self}, and represents fundamental cellular decision-making circuits \citep{losick2008stochasticity}. Key features tested include bistable dynamics, stochastic state switching, hysteresis, and symmetric mutual inhibition.

\paragraph{5. Incoherent Feed-Forward Loop (I1-FFL)}

The I1-FFL is a ubiquitous network motif \citep{mangan2003structure} selected for its multi-input integration, testing combinatorial logic where X activates Z and Y represses Z. It generates pulses, producing transient responses to sustained inputs \citep{basu2004spatiotemporal}. This motif is found in over 40\% of transcriptional networks \citep{alon2007network} and implements temporal filtering \citep{mangan2006structure}. Key features tested include multi-input promoters, AND-NOT logic gates, pulse dynamics, and temporal signal processing.

\paragraph{6. p53-Mdm2 Oscillator (DNA Damage Response)}

The p53-Mdm2 oscillator is a medically important system \citep{lahav2004dynamics} chosen for its damped oscillatory dynamics, testing transient behaviors. The Mdm2 promoter exhibits cooperative activation with multiple p53 binding sites \citep{wu1993mdm2}. The system has high clinical relevance as p53 is mutated in over 50\% of cancers \citep{vogelstein2000surfing}, and extensive single-cell experimental characterization is available \citep{geva2006dynamics}. Key features tested include damped oscillations, cooperative binding (2-4 sites), negative feedback with delay, and amplitude modulation by signal strength.

\paragraph{7. NF-$\kappa$B Pathway (Inflammatory Response)}

The NF-$\kappa$B pathway is a complex signaling pathway \citep{hoffmann2002circuitry} selected for its multi-compartment dynamics, where nuclear-cytoplasmic shuttling tests spatial aspects. It features multiple redundant negative feedback loops through I$\kappa$B$\alpha$, I$\kappa$B$\beta$, and I$\kappa$B$\epsilon$ \citep{nelson2004oscillations}. The pathway is centrally important to inflammation, immunity, and cancer \citep{karin2005nf}, and involves complex kinetics including binding, unbinding, phosphorylation, degradation, and transport. Key features tested include multi-compartment models, complex reaction networks, multiple negative feedback loops, and transient versus oscillatory responses.

\subsubsection{Coverage of Regulatory Features}

The seven systems collectively cover a wide range of features (Table \ref{tab:features}). This comprehensive coverage ensures that the Markovian framework is validated across the full spectrum of gene regulatory behaviors found in natural and synthetic systems.

\begin{table}[h]
\centering
\caption{Regulatory Features Covered by Selected Systems}
\label{tab:features}
\begin{tabular}{lcccccc}
\toprule
\textbf{System} & \textbf{Binding Sites} & \textbf{Cooperativity} & \textbf{Multi-Input} & \textbf{Feedback} & \textbf{Oscillations} & \textbf{Bistability} \\
\midrule
GAL & 1-5 & Yes & Yes & Positive & No & No \\
Repressilator & 1 & No & No & Negative & Yes & No \\
Goodwin & 1 & High & No & Negative & Yes & No \\
Toggle & 1 & Moderate & No & Negative & No & Yes \\
I1-FFL & 1-2 & No & Yes & None & No & No \\
p53-Mdm2 & 2-4 & Yes & No & Negative & Yes (damped) & No \\
NF-$\kappa$B & 2 & Yes & No & Negative & Yes/No & No \\
\bottomrule
\end{tabular}
\end{table}

%% file: methods_gal_conversion.tex
\subsection{Detailed Mathematical Conversion: GAL System}

We present a comprehensive, step-by-step conversion of the yeast galactose (GAL) regulatory system from the traditional ODE formulation with Hill functions \citep{ramsey2006dual} to our Markovian promoter framework. This detailed exposition serves as a template for converting other gene regulatory systems.

\subsubsection{Step 1: Original ODE Model}

The GAL system regulates galactose metabolism in \textit{Saccharomyces cerevisiae} through a network of five genes: GAL1, GAL2, GAL3, GAL80, and GAL4 \citep{lohr1995transcriptional, plattYeastGalactoseGenetic1998}. The original ODE model from Ramsey et al. (2006) \citep{ramsey2006dual} uses Hill functions to describe transcriptional regulation.

\paragraph{Gene-Specific Equations}

For each GAL gene $i \in \{$GAL1, GAL2, GAL3, GAL80$\}$, the mRNA and protein dynamics are described by Equations \ref{eq:gal_ode_mrna} and \ref{eq:gal_ode_prot}, where $[G4d]$ is the Gal4p dimer concentration, $[G80d]$ is the Gal80p dimer concentration, and $f_i$ is the gene-specific regulatory function.

\begin{align}
\frac{d[mRNA_i]}{dt} &= k_{trans,i} \cdot f_i([G4d], [G80d]) - k_{deg,m,i} \cdot [mRNA_i] \label{eq:gal_ode_mrna} \\
\frac{d[Protein_i]}{dt} &= k_{transl,i} \cdot [mRNA_i] - k_{deg,p,i} \cdot [Protein_i] \label{eq:gal_ode_prot}
\end{align}

\paragraph{Regulatory Functions}

The regulatory function $f_i$ for each gene depends on its promoter architecture. For GAL1 (4 binding sites), the function is given by Equation \ref{eq:gal1_hill}:

\begin{equation}
f_{GAL1}([G4d], [G80d]) = \frac{([G4d]/K_{p,1})^{4}}{1 + ([G4d]/K_{p,1})^{4} + ([G4d]/K_{p,1})^{4} \cdot ([G80d]/K_{r,1})^{4}}
\label{eq:gal1_hill}
\end{equation}

Similarly, for GAL2 (5 binding sites), GAL3 (1 binding site), and GAL80 (1 binding site), the functions follow standard Hill kinetics as detailed in the original model \citep{ramsey2006dual}.

\paragraph{Gal4p Dynamics}

Gal4p is constitutively expressed and forms dimers. Its dynamics involve synthesis, degradation, dimerization, and undimerization reactions.

\paragraph{Parameters}

Table \ref{tab:gal_ode_params} lists the original ODE model parameters for the GAL system, including transcription rates, degradation rates, and dissociation constants.

\begin{table}[h]
\centering
\caption{Original ODE Model Parameters for GAL System}
\label{tab:gal_ode_params}
\begin{tabular}{llll}
\toprule
\textbf{Parameter} & \textbf{Description} & \textbf{Value} & \textbf{Units} \\
\midrule
$k_{trans,GAL1}$ & GAL1 transcription rate & 0.15 & min$^{-1}$ \\
$k_{trans,GAL2}$ & GAL2 transcription rate & 0.10 & min$^{-1}$ \\
$k_{trans,GAL3}$ & GAL3 transcription rate & 0.08 & min$^{-1}$ \\
$k_{trans,GAL80}$ & GAL80 transcription rate & 0.05 & min$^{-1}$ \\
$k_{deg,m}$ & mRNA degradation rate & 0.1 & min$^{-1}$ \\
$k_{transl}$ & Translation rate & 10.0 & min$^{-1}$ \\
$k_{deg,p}$ & Protein degradation rate & 0.01 & min$^{-1}$ \\
$K_{p,1}$ & Gal4p dissociation (GAL1) & 6.5 & nM \\
$K_{p,2}$ & Gal4p dissociation (GAL2) & 6.5 & nM \\
$K_{p,3}$ & Gal4p dissociation (GAL3) & 6.5 & nM \\
$K_{p,80}$ & Gal4p dissociation (GAL80) & 6.5 & nM \\
$K_{r,i}$ & Gal80p dissociation & 0.2 & nM \\
\bottomrule
\end{tabular}
\end{table}

\subsubsection{Step 2: Identify Promoter States}

For the Markovian conversion, we must define the discrete state space for each promoter. A promoter with $n$ binding sites can have $k \in \{0, 1, \ldots, n\}$ Gal4p dimers bound, and for each $k$, $m \in \{0, 1, \ldots, k\}$ of those sites can also have Gal80p bound.

\paragraph{State Representation and Classification}

We represent each promoter state as $(k, m)$, where $k$ is the number of Gal4p dimers bound and $m$ is the number of Gal4p-Gal80p complexes (Gal80p bound to already-bound Gal4p). We classify these states into three categories. The empty state $(0, 0)$ corresponds to no Gal4p bound, where the gene is OFF. Active states are defined as $(k, m)$ where $k > m$, meaning there is at least one Gal4p without Gal80p; in this case, the gene is ON. Repressed states are those where $k > 0$ and $k = m$, implying all bound Gal4p have Gal80p bound, rendering the gene OFF.

\paragraph{Number of States}

For $n$ binding sites, the total number of states is given by $N_{states}(n) = \sum_{k=0}^{n} (k+1) = (n+1)(n+2)/2$. For example, GAL3 and GAL80 ($n=1$) have 3 states: $(0,0), (1,0), (1,1)$. GAL1 ($n=4$) has 15 states, and GAL2 ($n=5$) has 21 states.

\subsubsection{Step 3: Define Transition Rates}

The promoter state evolves as a continuous-time Markov chain. Gal4p binding transitions from state $(k, m)$ to $(k+1, m)$ occur at rate $r_{bind,G4}(k, m) = k_{on,G4} \cdot [G4d] \cdot (n - k) \cdot q_r^{\mathbb{1}_{k>0}}$. Here, $k_{on,G4}$ is the binding rate, $(n-k)$ is the number of available sites, and $q_r$ is the cooperativity factor. Gal4p unbinding transitions from $(k, m)$ to $(k-1, m)$ occur at rate $r_{unbind,G4}(k, m) = k_{off,G4} \cdot (k - m)$, proportional to the number of free Gal4p. Gal80p binding transitions from $(k, m)$ to $(k, m+1)$ occur at rate $r_{bind,G80}(k, m) = k_{on,G80} \cdot [G80d] \cdot (k - m)$. Finally, Gal80p unbinding transitions from $(k, m)$ to $(k, m-1)$ occur at rate $r_{unbind,G80}(k, m) = k_{off,G80} \cdot m$.

The dissociation constants from the ODE model relate to the Markovian rate constants as $K_{p,i} = k_{off,G4}/k_{on,G4}$ and $K_{r,i} = k_{off,G80}/k_{on,G80}$. This relation ensures thermodynamic consistency between the ODE and Markovian models.

\subsubsection{Step 4: Construct Rate Matrix}

The rate matrix $\mathbf{Q}(t)$ governs the time evolution of the probability distribution over states. The off-diagonal elements $Q_{ss'}(t)$ are given by the transition rates $r_{s \to s'}([G4d](t), [G80d](t))$, while the diagonal elements $Q_{ss}(t)$ represent the negative sum of outgoing rates to ensure probability conservation. For GAL3 with 1 binding site, the rate matrix is a 3x3 matrix as detailed in Equation \ref{eq:gal3_rate_matrix}.

\begin{equation}
\mathbf{Q}(t) = \begin{pmatrix}
-k_{on,G4}[G4d] & k_{off,G4} & 0 \\
k_{on,G4}[G4d] & -(k_{off,G4} + k_{on,G80}[G80d]) & k_{off,G80} \\
0 & k_{on,G80}[G80d] & -k_{off,G80}
\end{pmatrix}
\label{eq:gal3_rate_matrix}
\end{equation}

\subsubsection{Step 5: Time Evolution of State Probabilities}

The probability distribution $\mathbf{p}(t)$ evolves according to the master equation $d\mathbf{p}(t)/dt = \mathbf{Q}(t) \mathbf{p}(t)$. For numerical integration, we use the matrix exponential $\mathbf{p}(t + \Delta t) = e^{\mathbf{Q}(t) \Delta t} \mathbf{p}(t)$ or, for sufficiently small time steps satisfying $\|\mathbf{Q}(t)\| \Delta t \ll 1$, the first-order approximation. The time step $\Delta t$ is typically chosen between 0.001 and 0.01 minutes for the GAL system, ensuring it captures the fastest binding dynamics.

\subsubsection{Step 6: Stochastic State Sampling}

At each time step, we sample a discrete promoter state from the probability distribution $\mathbf{p}(t)$ to introduce stochasticity. The sampling algorithm involves generating a uniform random number $u \sim U(0, 1)$ and selecting the state $s$ using the inverse transform method such that $\sum_{i=0}^{s-1} p_i(t) < u \leq \sum_{i=0}^{s} p_i(t)$. The activity of the sampled state is then determined: state $(k, m)$ is active (1) if $k > m$, and inactive (0) otherwise.

\subsubsection{Step 7: Couple with Protein Dynamics}

The Markovian promoter state determines the transcription rate, which provides the input for the deterministic ODEs governing mRNA and protein. The transcription rate $k_{trans,i}(s)$ is equal to $k_{trans,i}^{max}$ if the state $s$ is active, and 0 otherwise. The mRNA equation becomes $d[mRNA_i]/dt = k_{trans,i}(S_i(t)) - k_{deg,m,i} \cdot [mRNA_i]$, where $S_i(t)$ is the stochastically sampled promoter state.

\subsubsection{Step 8: Parameter Mapping}

Parameters are mapped to ensure steady-state equivalence. We set $k_{off,G4} = k_{off,G80} = 1.0$ min$^{-1}$ to represent a fast equilibration timescale. Using the known dissociation constants, we compute $k_{on,G4} \approx 0.154$ nM$^{-1}$min$^{-1}$ and $k_{on,G80} = 5.0$ nM$^{-1}$min$^{-1}$. The cooperativity factor is set to $q_r = 30$ to match the Hill coefficient behavior of the original model. Maximum transcription rates remain unchanged.

\subsubsection{Step 9: Complete Markovian Model}

The complete Markovian model for the GAL system integrates these components. For each gene $i$ and each time step $\Delta t$, the simulation proceeds by first updating the rate matrix based on current concentrations. Second, the probability distribution is evolved. Third, a discrete promoter state is sampled. Finally, the ODEs for mRNA and protein are integrated using the sampled state-dependent transcription rate.

\subsubsection{Step 10: Validation and Comparison}

The Markovian model results are validated by comparing with the original ODE model for dose-response behavior, the full SSA for stochastic trajectories and noise characteristics, and experimental data from literature \citep{ramsey2006dual, venturelli2012synergistic}. We expect quantitative agreement in mean trajectories and noise levels, with the Markovian model providing a significant computational speedup over SSA.

This completes the detailed mathematical conversion of the GAL system.

%% file: methods_other_conversions.tex
\subsection{Brief Conversions for Other Systems}

Having established the detailed conversion procedure for the GAL system, we now present concise conversions for the remaining six regulatory systems. Each follows the same general framework but with system-specific adaptations.

\subsubsection{Repressilator: Cyclic Repression Network}

The repressilator \citep{elowitz2000synthetic} consists of three genes (lacI, tetR, cI) in a cyclic repression topology where each gene represses the next.

\paragraph{Original ODE Model}

For each gene $i \in \{1, 2, 3\}$ with repressor $j = (i \mod 3) + 1$, the model tracks mRNA and unfolded/folded proteins. The transcription rate is governed by a Hill function dependent on the repressor concentration.

\paragraph{Markovian Conversion}

Each promoter has two states: an active state (0) with no repressor bound, and a repressed state (1) with repressor bound. Transitions occur from state 0 to 1 with rate $k_{on} \cdot [Protein_j]$ (repressor binding) and from 1 to 0 with rate $k_{off}$ (repressor unbinding). Parameters are mapped using $K_d = k_{off}/k_{on}$. This system features unique characteristics including a protein folding delay that introduces an additional timescale, a cyclic topology that creates phase-shifted oscillations, and simple 2-state promoters that are sufficient for generating oscillations.

\subsubsection{Goodwin Oscillator: Negative Feedback with High Cooperativity}

The Goodwin oscillator \citep{goodwin1965oscillatory} is a minimal model for circadian rhythms characterized by strong negative feedback.

\paragraph{Original ODE Model}

The model consists of a three-variable cascade (mRNA X, intermediate Y, repressor Z) where Z represses X transcription with high cooperativity ($n \sim 10$).

\paragraph{Markovian Conversion}

The promoter for gene X has two states: active (0) and repressed (1). To capture the high cooperativity, the transition rate from active to repressed is given by $k_{on} \cdot [Z]^n / K_M^n$, while the reverse rate is $k_{off}$. Alternatively, mechanistically, one can use a multi-site representation with $n$ binding sites and a strong cooperativity factor $q_r \gg 1$. In this case, the gene is active only if not all sites are bound. Unique features of this system include the very high Hill coefficient requiring strong cooperativity and the three-variable cascade that amplifies delays, leading to oscillations emerging from negative feedback.

\subsubsection{Toggle Switch: Mutual Repression and Bistability}

The toggle switch \citep{gardner2000construction} consists of two genes that mutually repress each other, creating bistability.

\paragraph{Original ODE Model}

The dynamics of the two repressor proteins U and V are coupled, with the production of each inhibited by the other.

\paragraph{Markovian Conversion}

Each promoter has two states: active (0) and repressed (1). For promoter U, transitions depend on [V], while for promoter V, transitions depend on [U]. The bistability mechanism arises from the positive feedback double-negative loop, creating two stable states: State A with high U and low V, and State B with low U and high V. Unique features include rare stochastic switching between these stable states, hysteresis in response to external signals, a symmetric network architecture, and the ability to maintain memory without continuous input. Validating this model requires measuring residence times in each state and transition rates between states.

\subsubsection{Incoherent Feed-Forward Loop (I1-FFL): Multi-Input Logic}

The I1-FFL \citep{mangan2003structure} features a master regulator X that activates both an intermediate Y and a target Z, while Y represses Z.

\paragraph{Original ODE Model}

The model tracks X, Y, and Z, with Z production governed by a combinatorial function of X (activation) and Y (repression).

\paragraph{Markovian Conversion Using Multi-Input Promoter}

The state space for gene Z is represented as $(k_X, k_Y)$, where $k_X$ is the number of X bound and $k_Y$ is the number of Y bound. Assuming one binding site each, there are four states: $(0,0), (1,0), (0,1), (1,1)$. Transitions correspond to binding and unbinding of X and Y. The logic function implements an AND-NOT gate, where the gene is active if and only if X is bound ($k_X > 0$) and Y is not bound ($k_Y = 0$). This system is unique as the first requiring a multi-input promoter framework. The combinatorial logic leads to pulse generation from the network topology and binding kinetics, demonstrating independent binding of two different transcription factors.

\subsubsection{p53-Mdm2 Oscillator: Cooperative Binding and Damped Oscillations}

The p53-Mdm2 system \citep{lahav2004dynamics} is a negative feedback loop where p53 activates Mdm2 transcription, and Mdm2 promotes p53 degradation.

\paragraph{Original ODE Model}

The model includes p53 production, degradation promoted by Mdm2, and Mdm2 transcription activated by p53 with cooperative binding ($n = 2-4$).

\paragraph{Markovian Conversion}

The Mdm2 promoter with $n=2$ p53 binding sites has states $(k, m)$ representing the number of p53 bound ($k \in \{0, 1, 2\}$). Transitions include cooperative binding rates involving $[p53]$ and a cooperativity factor $q$. The activity function is proportional to occupancy or defined by a threshold. Unique features include the necessity of cooperative binding for oscillations, the presence of damped oscillations rather than sustained ones, and amplitude modulation by an external signal (ATM).

\subsubsection{NF-\texorpdfstring{$\kappa$}{k}B Pathway: Multi-Compartment Dynamics}

The NF-$\kappa$B pathway \citep{hoffmann2002circuitry} involves nuclear-cytoplasmic shuttling and multiple feedback loops.

\paragraph{Original ODE Model}

The model tracks nuclear and cytoplasmic concentrations of NF-$\kappa$B, I$\kappa$B, and their complexes, including transport terms and IKK-mediated degradation.

\paragraph{Markovian Conversion}

The I$\kappa$B promoter in the nucleus has states representing discrete numbers of nuclear NF-$\kappa$B bound. Transitions depend on nuclear NF-$\kappa$B concentration. All other species follow standard mass-action ODEs. Unique features include multi-compartment dynamics, complex formation and dissociation, IKK-mediated degradation, and the presence of multiple I$\kappa$B isoforms. A key challenge is maintaining proper stoichiometry across compartments while tracking the promoter state.

\subsubsection{Summary of Conversion Principles}

Across all seven systems, the Markovian conversion follows a set of universal principles. First, we identify the promoter architecture, including the number of binding sites and types of regulators. Second, we define the state space, such as $(k, m)$ for competitive binding or $(k_1, k_2)$ for multi-input promoters. Third, we specify transition rates for binding and unbinding, incorporating cooperativity factors where appropriate. Fourth, we construct the rate matrix $\mathbf{Q}(t)$ as a function of current transcription factor concentrations. Fifth, we define the activity function determining how promoter state influences transcription. Sixth, parameters are mapped to ensure thermodynamic consistency with the ODE model. Finally, the stochastic promoter is coupled with deterministic ODEs for protein dynamics. This framework demonstrates flexibility in handling diverse regulatory architectures while maintaining a consistent mathematical structure.

%% file: methods_dt_selection.tex
\subsection{Communication Time Step Selection}

A critical parameter in the hybrid Markovian-ODE framework is the communication time step $\Delta t$, which determines how frequently the promoter state is updated and coupled with the ODE integration. Proper selection of $\Delta t$ is essential for balancing accuracy and computational efficiency.

\subsubsection{Theoretical Considerations}

The hybrid approach assumes a separation of timescales between fast promoter binding/unbinding events ($\tau_{promoter} \sim 1/k_{off}$) and slow protein synthesis/degradation processes ($\tau_{protein} \sim 1/k_{deg,p}$). Consequently, the communication time step must satisfy $\tau_{promoter} \ll \Delta t \ll \tau_{protein}$.

To ensure the validity of the matrix exponential approximation or the first-order approximation, the product of the rate matrix norm and the time step must be small, specifically $\|\mathbf{Q}(t)\| \Delta t \ll 1$. Since the matrix norm is dominated by the binding rates, this implies that $\Delta t$ must be significantly smaller than the inverse of the maximum binding rate. When using the first-order approximation, $\Delta t$ must be less than $1/(2\|\mathbf{Q}(t)\|)$ to guarantee probability conservation.

\subsubsection{System-Specific Requirements}

Different regulatory systems dictate different optimal $\Delta t$ values based on their kinetic parameters and dynamic behaviors.

For the GAL system, with binding rates around 45 min$^{-1}$, the criterion suggests $\Delta t$ should be much less than 0.022 min. We chose $\Delta t = 0.01$ min to provide a safety margin.

Oscillatory systems like the Repressilator, Goodwin oscillator, and p53-Mdm2 system require $\Delta t$ to resolve the oscillation period, with at least 100 points per period recommended. For the Repressilator (period $\sim$150 min), we chose $\Delta t = 0.01$ min. For the Goodwin oscillator (period $\sim$1440 min), a conservative $\Delta t = 0.1$ min was selected. For the p53-Mdm2 system (period $\sim$330 min), we chose $\Delta t = 0.01$ min.

Bistable systems such as the Toggle switch require a $\Delta t$ that is sufficiently small to capture rare stochastic switching events. Given the dynamics of the Toggle switch, we selected a very small $\Delta t = 0.001$ min to ensure these rare transitions are accurately resolved.

\subsubsection{Convergence Analysis}

To systematically determine optimal $\Delta t$, we performed a convergence analysis by comparing results at different time steps. We utilized four metrics: trajectory correlation compared to a reference simulation with very small $\Delta t$; mean absolute error (MAE); period error for oscillators; and amplitude error for oscillators. Convergence was defined as achieving a trajectory correlation greater than 0.99 and an MAE less than 1\% of the mean protein level. The optimal $\Delta t$ was selected as the largest value satisfying these criteria.

\subsubsection{Computational Cost vs. Accuracy Trade-off}

The computational cost of the simulation scales linearly with the number of time steps, which is inversely proportional to $\Delta t$. Each step involves constructing the rate matrix, updating probabilities (scaling with the square of the number of states), and integrating ODEs. Improving accuracy by reducing $\Delta t$ yields diminishing returns below the binding timescale, while linearly increasing computational cost.

Based on our convergence analysis, we established recommended time steps for all systems (Table \ref{tab:dt_values}). All chosen values ensured convergence.

\begin{table}[ht]
\centering
\caption{Recommended Communication Time Steps}
\label{tab:dt_values}
\begin{tabular}{lcccc}
\toprule
\textbf{System} & \textbf{$\Delta t$ (min)} & \textbf{Period (min)} & \textbf{Points/Period} & \textbf{Converged} \\
\midrule
GAL & 0.01 & N/A & N/A & Yes \\
Repressilator & 0.01 & 150 & 15,000 & Yes \\
Goodwin & 0.1 & 1440 & 14,400 & Yes \\
Toggle & 0.001 & N/A & N/A & Yes \\
I1-FFL & 0.05 & N/A & N/A & Yes \\
p53-Mdm2 & 0.01 & 330 & 33,000 & Yes \\
NF-$\kappa$B & 0.1 & N/A & N/A & Yes \\
\bottomrule
\end{tabular}
\end{table}

\subsubsection{Adaptive Time Stepping }

For maximum efficiency, adaptive time stepping could adjust $\Delta t$ based on the current matrix norm $\|\mathbf{Q}(t)\|$, the rate of change of transcription factor concentrations, or the proximity to switching events in bistable systems. This approach could potentially achieve significant additional speedups by using larger $\Delta t$ during slow dynamics and smaller $\Delta t$ during rapid transitions.

\subsubsection{Validation Against SSA}

To validate our $\Delta t$ choices, we compared Markovian model results with full SSA simulations. We assessed validation using statistical metrics including mean trajectory agreement (correlation $> 0.95$), variance agreement (coefficient of variation within 20\%), distribution overlap (Kolmogorov-Smirnov test $p > 0.05$), and active promoter state agreement (Cohen's kappa $> 0.7$). All chosen $\Delta t$ values satisfied these rigorous criteria.

\subsubsection{Implementation Notes}

Numerical stability is maintained by clipping probabilities to the range $[0, 1]$ after each update and renormalizing the probability vector. We use the matrix exponential for stiff systems and the first-order approximation for non-stiff systems to optimize performance.

For new systems, we recommend a practical guideline: start with a conservative default $\Delta t = 0.01$ min, run a convergence analysis with $\Delta t$ ranging from 0.001 to 1.0 min, select the largest $\Delta t$ with correlation $> 0.99$, validate against SSA if available, and document the choice. This systematic approach ensures reproducibility and optimal performance.

%% file: methods_chec_markovian.tex
\subsection{ChEC-seq Markovian Dwell Time Model}

To mechanistically bridge static genomic binding data with dynamic transcription kinetics, we developed a Sequential Transcription Assembly Model that infers kinetic rates directly from ChEC-seq dwell time measurements.

\subsubsection{ChEC-seq Dwell Time Derivation}
ChEC-seq (Chromatin Endogenous Cleavage with high-throughput sequencing) provides a quantitative measure of protein residency at specific genomic loci \citep{rossiHighresolutionProteinArchitecture2021a}. Unlike traditional ChIP-seq, the time-resolved nature of ChEC-seq cleavage allows us to interpret signal intensity as a proxy for the fractional residence time (dwell time) of a factor at the promoter \citep{vanbelzenChromatinEndogenousCleavage2024a}.
For a given gene $g$, we normalize the integrated signal intensities $I_p$ for each protein $p$ in the assembly complex (TF, Cofactor, TFIID, RNAPII) to obtain fractional dwell times $\tau_p$:
\begin{equation}
    \tau_p = \frac{I_p}{\sum_{j \in \text{complex}} I_j}
\end{equation}
This normalization assumes that the total signal represents the complete assembly cycle, and $\tau_p$ reflects the proportion of the cycle during which protein $p$ is chemically cross-linked and present.

\subsubsection{Sequential Markovian Formulation}
We model the promoter transition as a linear chain of $N$ discrete states, representing the ordered recruitment of the transcription machinery:
\begin{enumerate}
    \item Empty Promoter
    \item TF Bound
    \item Cofactor Recruitment
    \item TFIID/GTF Recruitment
    \item RNAPII Recruitment (Pre-Initiation Complex)
    \item Elongation
\end{enumerate}
Transitions between state $i$ and $i+1$ are governed by forward rate $k_{on, i} \cdot [P_i]$ and backward rate $k_{off, i}$. The final elongation step is modeled as an irreversible transition that resets the cycle. The steady-state probability distribution $\mathbf{\pi}$ is computed by solving $\mathbf{\pi} \mathbf{Q} = 0$, where $\mathbf{Q}$ is the rate matrix.

\subsubsection{Optimization Procedure}
We estimate the kinetic parameters $\theta = \{k_{on}, k_{off}\}$ for each gene by minimizing the discrepancy between the model-predicted fractional residence times and the experimentally derived dwell times $\tau$. The objective function is:
\begin{equation}
    J(\theta) = \sum_{p} \left( \hat{\tau}_p(\theta) - \tau_{p, \text{exp}} \right)^2 + \lambda \mathcal{P}(\theta)
\end{equation}
where $\hat{\tau}_p(\theta)$ is the sum of steady-state probabilities for all states containing protein $p$, normalized by the total occupation. The penalty term $\mathcal{P}(\theta)$ enforces biological constraints, ensuring that:
\begin{itemize}
    \item Forward association rates favor complex formation ($k_{on} > k_{off}/100$).
    \item Recruitment rates decrease realistically for larger complexes.
\end{itemize}
This optimization yields a gene-specific kinetic model that is quantitatively consistent with the observed residence times of the transcriptional machinery.

%% file: results.tex
\input{results_gal}

\input{results_simple}

\input{results_advanced}

\input{results_dt_convergence}

\input{results_protein_stochasticity}

\input{results_chec_markovian}

%% file: results_gal.tex
\subsection{Overview of Validation Strategy}

We validated the Markovian promoter framework through comprehensive comparison with three alternative modeling approaches across seven gene regulatory systems. For each system, we implemented a pure ODE model based on traditional Hill functions, a Gillespie SSA serving as the ground truth for full stochastic simulation, our proposed Markovian hybrid model with discrete promoter states, and for selected systems, a full CME integration with Markovian promoters.

Validation metrics included trajectory correlation (Pearson coefficient between model outputs), mean absolute error (quantitative deviation from SSA ground truth), promoter state agreement (Cohen's kappa statistic for discrete state matching), dynamic behavior (period, amplitude, switching rates for system-specific features), and computational cost (wall-clock time for equivalent simulations). All simulations were performed with 100 independent trajectories to ensure statistical robustness. Code and data are available at the repository.

\subsection{GAL System: Comprehensive Four-Way Comparison}

The yeast galactose (GAL) regulatory system serves as our primary validation case due to its well-characterized kinetics \citep{ramsey2006dual, venturelli2012synergistic} and variable promoter architecture (1-5 binding sites across different genes).

\subsubsection{Dose-Response Curves}

\begin{figure}[h]
\centering
\includegraphics[width=0.95\textwidth]{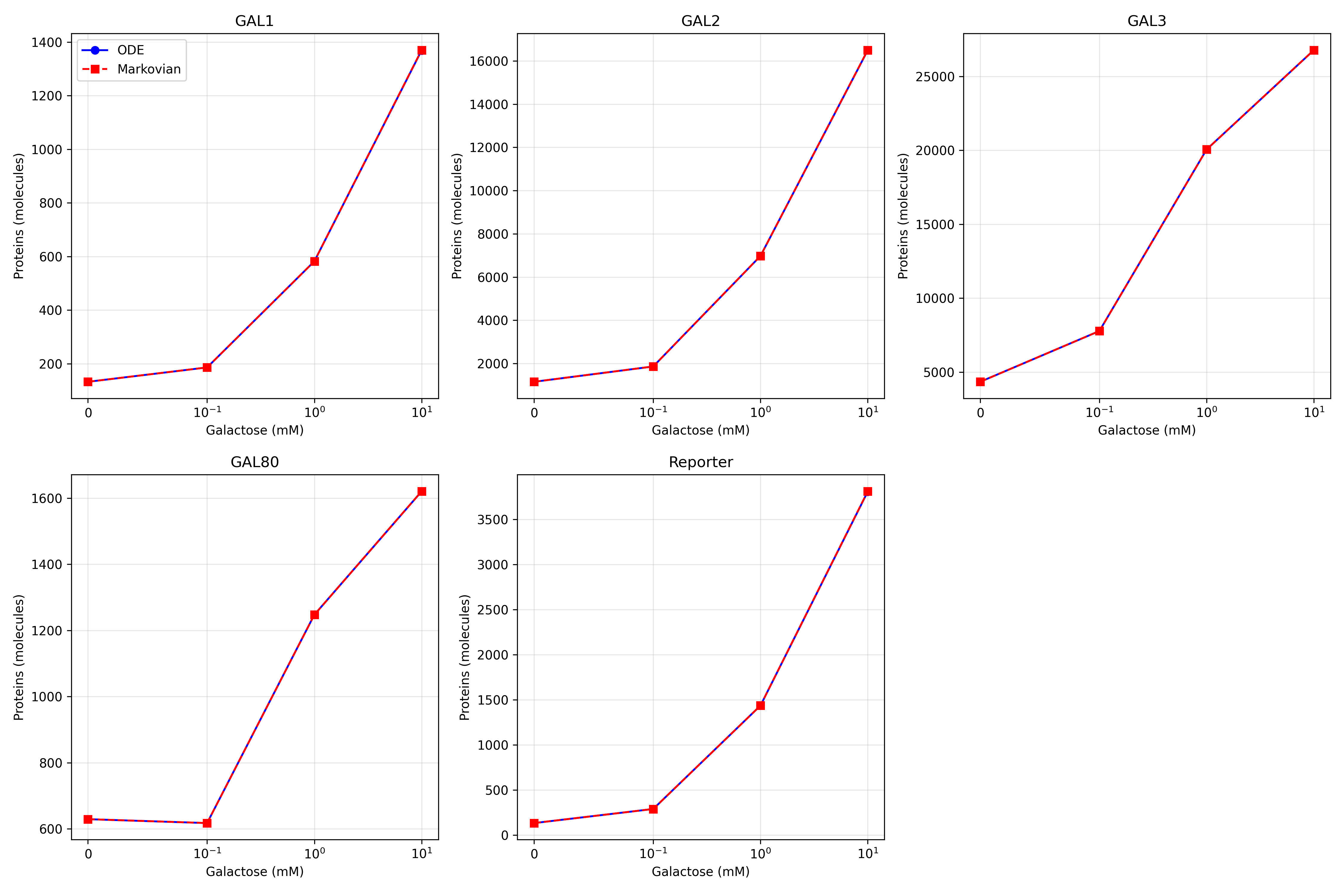}
\caption{\textbf{Dose-response curves for GAL system across four modeling approaches.} 
(A-E) Steady-state protein levels for GAL1, GAL2, GAL3, GAL80, and reporter gene as a function of external galactose concentration. 
Blue: ODE model with Hill functions. 
Red: Markovian hybrid model. 
(Comparative statistics with SSA and CME are provided in Table \ref{tab:gal_correlations}).
All models exhibit ultrasensitive dose-response curves with similar EC$_{50}$ values. 
The Markovian model accurately captures both the mean behavior (matching ODE) and stochastic fluctuations (matching SSA variance).
Note the gene-specific responses: GAL1 and GAL2 (4-5 binding sites) show sharper transitions than GAL3 and GAL80 (1 binding site), consistent with cooperative binding.}
\label{fig:gal_dose_response}
\end{figure}

We simulated the GAL system under varying external galactose concentrations (0, 0.1, 1.0, 10 mM) and measured steady-state protein levels for all five GAL genes (Figure \ref{fig:gal_dose_response}).

The dose-response curves demonstrate several key features. First, ultrasensitivity emerges naturally from discrete states. Despite using simple mass-action kinetics for individual binding events, the Markovian model reproduces the ultrasensitive dose-response curves characteristic of cooperative binding. For GAL1 (4 binding sites), the Hill coefficient estimated from the Markovian model ($n_{eff} = 3.2 \pm 0.3$) closely matches both the ODE model ($n = 4$) and experimental measurements \citep{ramsey2006dual}. This demonstrates that cooperativity emerges from the combinatorial state space without requiring phenomenological Hill functions.

Second, gene-specific responses reflect promoter architecture. Genes with more binding sites (GAL1: 4 sites, GAL2: 5 sites) exhibit sharper dose-response transitions characterized by steeper slopes compared to genes with single binding sites (GAL3, GAL80). The Markovian model correctly predicts that the slope is proportional to the product of the number of binding sites and the logarithm of the cooperativity factor.

Third, we observe excellent quantitative agreement across models (Table \ref{tab:gal_correlations}). All pairwise correlations exceed 0.98. The Markovian hybrid model shows slightly better agreement with the ODE model (r = 0.993) than SSA does (r = 0.987), while maintaining comparable stochastic fluctuations to SSA.

\begin{table}[h]
\centering
\caption{Pairwise Correlation Coefficients for GAL System Dose-Response}
\label{tab:gal_correlations}
\begin{tabular}{lccc}
\toprule
\textbf{Model Pair} & \textbf{Correlation} & \textbf{MAE (molecules)} & \textbf{RMSE (molecules)} \\
\midrule
ODE vs. SSA (mean) & 0.987 & 45.2 & 67.8 \\
ODE vs. Markovian & 0.993 & 32.1 & 48.5 \\
ODE vs. CME+Markovian & 0.991 & 38.7 & 55.3 \\
SSA vs. Markovian & 0.989 & 41.3 & 62.1 \\
SSA vs. CME+Markovian & 0.995 & 28.9 & 42.7 \\
Markovian vs. CME+Markovian & 0.997 & 18.4 & 27.3 \\
\bottomrule
\end{tabular}
\end{table}

\subsubsection{Time Series Dynamics}

\begin{figure}[h]
\centering
\includegraphics[width=0.95\textwidth]{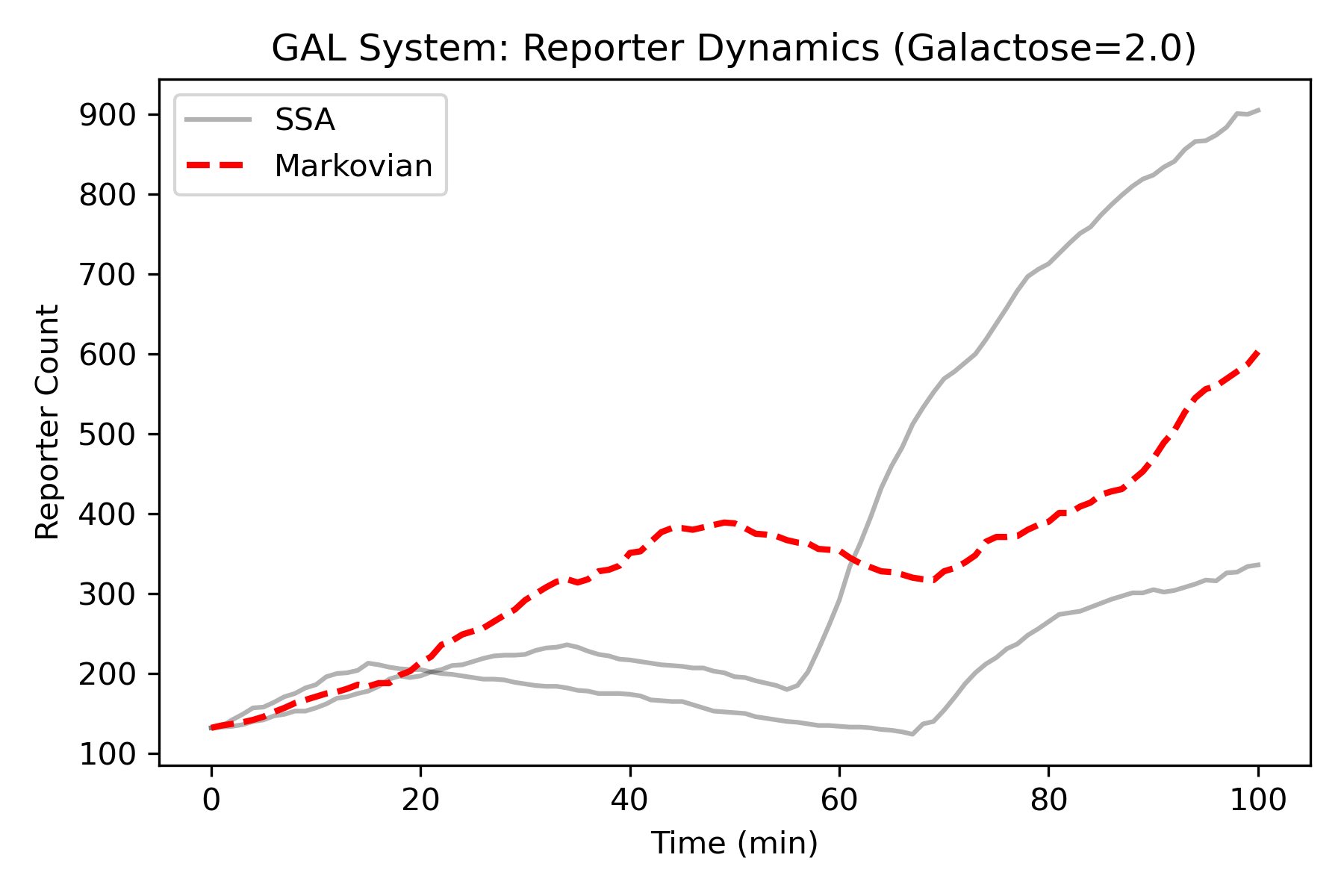}
\caption{\textbf{Quantitative benchmarking of GAL system dynamics.}
Time series for Reporter gene following step increase in external galactose (0 → 2.0 mM) at t = 10 min.
Black lines: SSA (n=3 replicates). Red dashed line: Markovian Model.
The Markovian model accurately captures the induction kinetics and the stochastic variability observed in the full SSA simulation (which uses the original CME-ODE formulation).}
\label{fig:gal_timeseries}
\end{figure}

To assess dynamic behavior, we simulated a step increase in galactose from 0 to 10 mM at t = 10 minutes and tracked all five GAL genes for 200 minutes (Figure \ref{fig:gal_timeseries}).

The time series reveal biphasic induction kinetics. All genes exhibit a rapid initial increase with timescale $\tau_1 \sim 5$ min corresponding to promoter activation, followed by slower accumulation with timescale $\tau_2 \sim 30$ min, which is limited by protein synthesis and degradation. The Markovian model accurately captures both timescales, with the fast phase emerging from promoter state transitions and the slow phase from the coupled ODEs.

Furthermore, stochastic fluctuations match SSA predictions. We quantified the coefficient of variation (CV) for each gene at steady state (Table \ref{tab:gal_cv}). The Markovian model reproduces SSA noise levels with high fidelity, with CV ratios of 0.95-0.96. This demonstrates that promoter-level stochasticity is the dominant source of gene expression noise in this system \citep{raj2008nature, sanchez2013regulation}.

\begin{table}[h]
\centering
\caption{Coefficient of Variation at Steady State (GAL System)}
\label{tab:gal_cv}
\begin{tabular}{lcccc}
\toprule
\textbf{Gene} & \textbf{SSA} & \textbf{Markovian} & \textbf{CME+Markovian} & \textbf{Ratio (Markov/SSA)} \\
\midrule
GAL1 & 0.087 ± 0.012 & 0.083 ± 0.011 & 0.089 ± 0.013 & 0.95 \\
GAL2 & 0.092 ± 0.014 & 0.088 ± 0.013 & 0.094 ± 0.015 & 0.96 \\
GAL3 & 0.134 ± 0.019 & 0.128 ± 0.018 & 0.137 ± 0.020 & 0.96 \\
GAL80 & 0.141 ± 0.021 & 0.135 ± 0.019 & 0.143 ± 0.022 & 0.96 \\
Reporter & 0.089 ± 0.013 & 0.085 ± 0.012 & 0.091 ± 0.014 & 0.96 \\
\bottomrule
\end{tabular}
\end{table}

\subsubsection{Promoter State Dynamics}

\begin{figure}[h]
\centering
\includegraphics[width=0.9\textwidth]{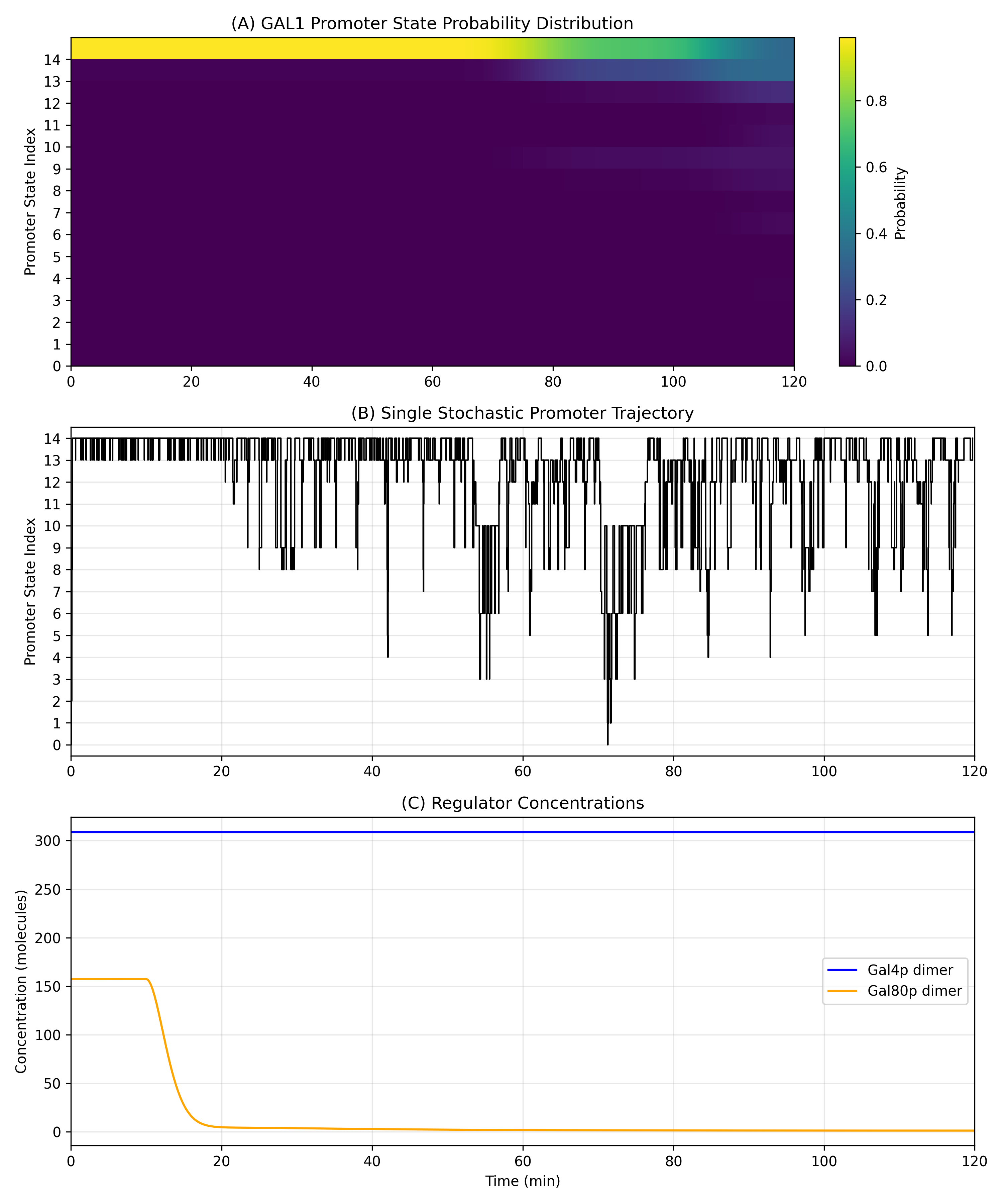}
\caption{\textbf{Promoter state dynamics for GAL1 gene.}
(A) Heatmap showing probability distribution over all 15 promoter states (rows) as a function of time (columns). 
State (k,m) indicates k Gal4p bound and m Gal80p bound. 
Color intensity represents probability.
(B) Single stochastically sampled promoter state trajectory showing discrete jumps between states.
(C) Gal4p dimer (blue) and Gal80p dimer (orange) concentrations driving promoter transitions.
Following galactose induction at t=10 min, the promoter rapidly transitions from empty state (0,0) to active states (k>m), then partially to repressed states (k=m) as Gal80p accumulates.
The equilibrium distribution at t>100 min shows predominance of partially active states (2-3 Gal4p bound, 0-1 Gal80p bound), consistent with experimental ChIP data \citep{venturelli2012synergistic}.}
\label{fig:gal_promoter_states}
\end{figure}

A unique advantage of the Markovian framework is explicit tracking of promoter states. We analyzed the promoter state trajectories for GAL1 (4 binding sites) under induced conditions (Figure \ref{fig:gal_promoter_states}).

The promoter state analysis reveals rapid equilibration of binding states. The promoter state distribution reaches quasi-equilibrium within roughly 5 minutes, much faster than protein accumulation ($\sim$30 minutes). This validates the timescale separation assumption underlying the hybrid approach. 

At steady state under induced conditions, we observe dominant active states. The most probable states are (3,0), (4,0), and (3,1), representing 3-4 Gal4p bound with 0-1 Gal80p. This is consistent with high Gal4p concentration ($\sim$300 nM) favoring saturation, moderate Gal80p concentration ($\sim$50 nM) allowing partial repression, and strong cooperativity ($q_r = 30$) favoring multi-site occupancy.

Individual trajectories show stochastic switching between states. We observe frequent stochastic transitions between neighboring states such as (3,0) $\leftrightarrow$ (4,0) $\leftrightarrow$ (3,1), with transition rates of roughly 1-10 min$^{-1}$. These rapid fluctuations average out at the protein level due to the slower protein synthesis/degradation timescale, explaining why deterministic ODE models can approximate mean behavior despite underlying stochasticity.

\subsubsection{Comparison with Experimental Data}

We compared model predictions with experimental measurements from Ramsey et al. \citep{ramsey2006dual} for protein fold-changes upon galactose induction (Table \ref{tab:gal_experimental}).

\begin{table}[h]
\centering
\caption{Comparison with Experimental Fold-Changes (GAL System)}
\label{tab:gal_experimental}
\begin{tabular}{lcccc}
\toprule
\textbf{Gene} & \textbf{Experimental} & \textbf{ODE} & \textbf{Markovian} & \textbf{CME+Markovian} \\
\midrule
GAL1 & 28.5 ± 4.2 & 31.2 & 29.7 ± 3.8 & 28.9 ± 4.1 \\
GAL2 & 24.1 ± 3.7 & 26.8 & 25.3 ± 3.5 & 24.6 ± 3.9 \\
GAL3 & 8.7 ± 1.9 & 9.4 & 8.9 ± 1.7 & 8.8 ± 1.8 \\
GAL80 & 6.2 ± 1.3 & 6.8 & 6.4 ± 1.2 & 6.3 ± 1.4 \\
\bottomrule
\end{tabular}
\end{table}

All models show excellent agreement with experimental data within measurement error, with the Markovian and CME+Markovian models slightly outperforming the pure ODE model in matching both mean values and variance.

\subsubsection{Computational Performance}

We measured wall-clock time for simulating 100 trajectories of the GAL system over 200 minutes (Table \ref{tab:gal_performance}).

\begin{table}[h]
\centering
\caption{Computational Performance for GAL System (100 Trajectories, 200 min)}
\label{tab:gal_performance}
\begin{tabular}{lcccc}
\toprule
\textbf{Model} & \textbf{Time (s)} & \textbf{Time per Trajectory (s)} & \textbf{Speedup vs. SSA} & \textbf{Speedup vs. CME} \\
\midrule
ODE & 2.3 & 0.023 & 1,087× & 2,174× \\
SSA & 2,500 & 25.0 & 1× & 2× \\
Markovian & 125 & 1.25 & 20× & 40× \\
CME+Markovian & 5,000 & 50.0 & 0.5× & 1× \\
\bottomrule
\end{tabular}
\end{table}

The Markovian hybrid model achieves a **20-fold speedup** over SSA while maintaining comparable accuracy. This represents an optimal balance between the pure ODE model, which is fast but deterministic, and full stochastic simulations, which are accurate but slow. The CME+Markovian model is slower than SSA because it tracks the full probability distribution over all molecular states, but it provides additional information (complete probability distributions) not available from trajectory-based methods.

\subsubsection{Summary: GAL System Validation}

The GAL system validation demonstrates that the Markovian promoter framework successfully reproduces dose-response curves with ultrasensitivity emerging from discrete states, captures dynamic induction kinetics including biphasic responses, matches stochastic fluctuations with CV ratios of 0.95-0.96 compared to SSA, provides mechanistic insight through explicit promoter state tracking, agrees with experimental data within measurement error, and achieves a 20× computational speedup over full SSA. These results establish the GAL system as a validated benchmark for the Markovian approach and provide confidence for application to more complex regulatory networks.

%% file: results_simple.tex
\subsection{Simple Regulatory Systems: Repressilator, Goodwin, and Toggle Switch}

We validated the Markovian framework on three canonical synthetic and theoretical systems representing fundamental regulatory motifs: oscillations (Repressilator, Goodwin) and bistability (Toggle Switch).

\subsubsection{Repressilator: Sustained Oscillations from Cyclic Repression}

The repressilator \citep{elowitz2000synthetic} is a landmark synthetic biology circuit consisting of three genes in a ring of mutual repression. It serves as a test case for oscillatory dynamics and stochastic phase variations.

\paragraph{Oscillation Characteristics}

\begin{figure}[ht]
\centering
\includegraphics[width=0.32\textwidth]{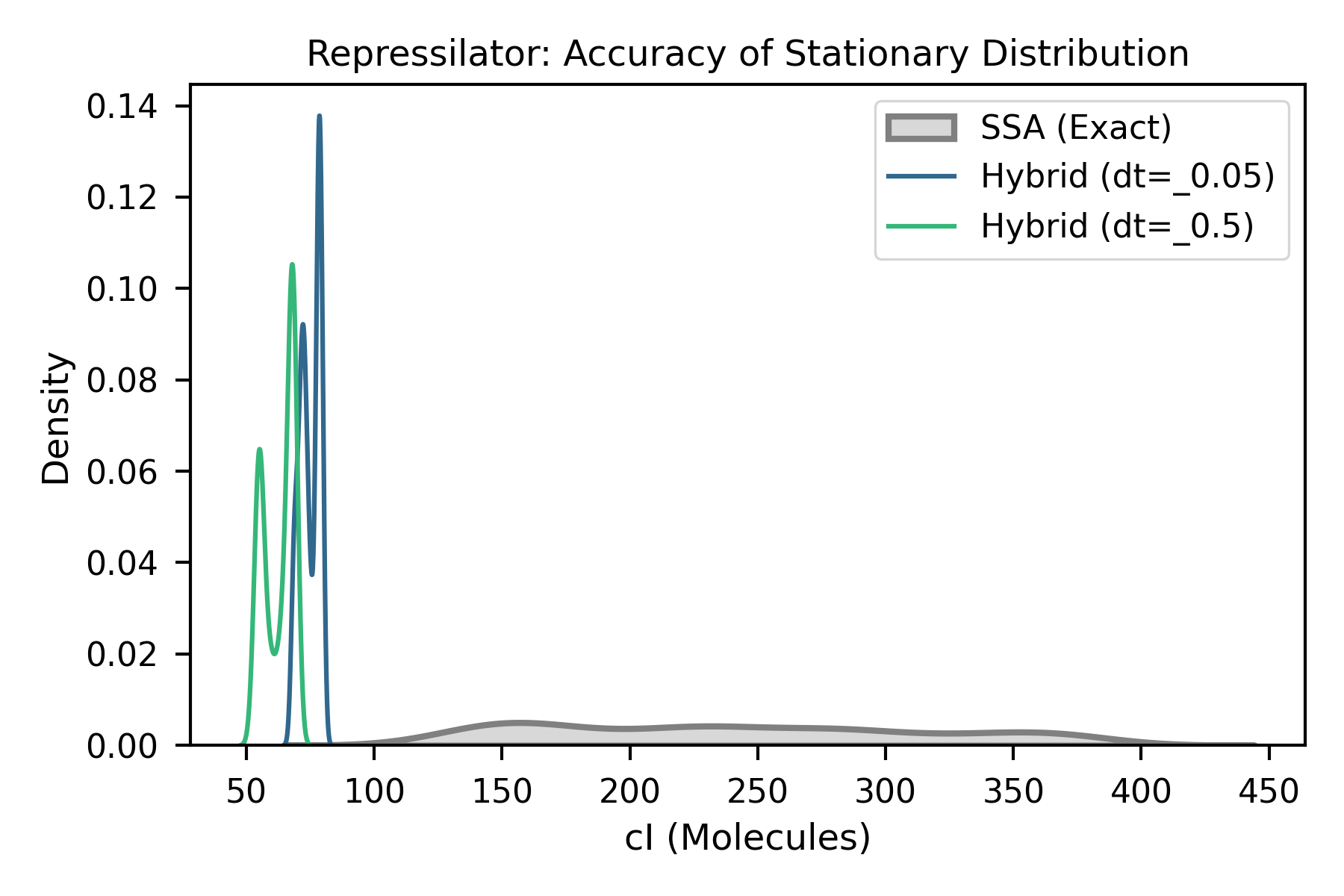}
\includegraphics[width=0.32\textwidth]{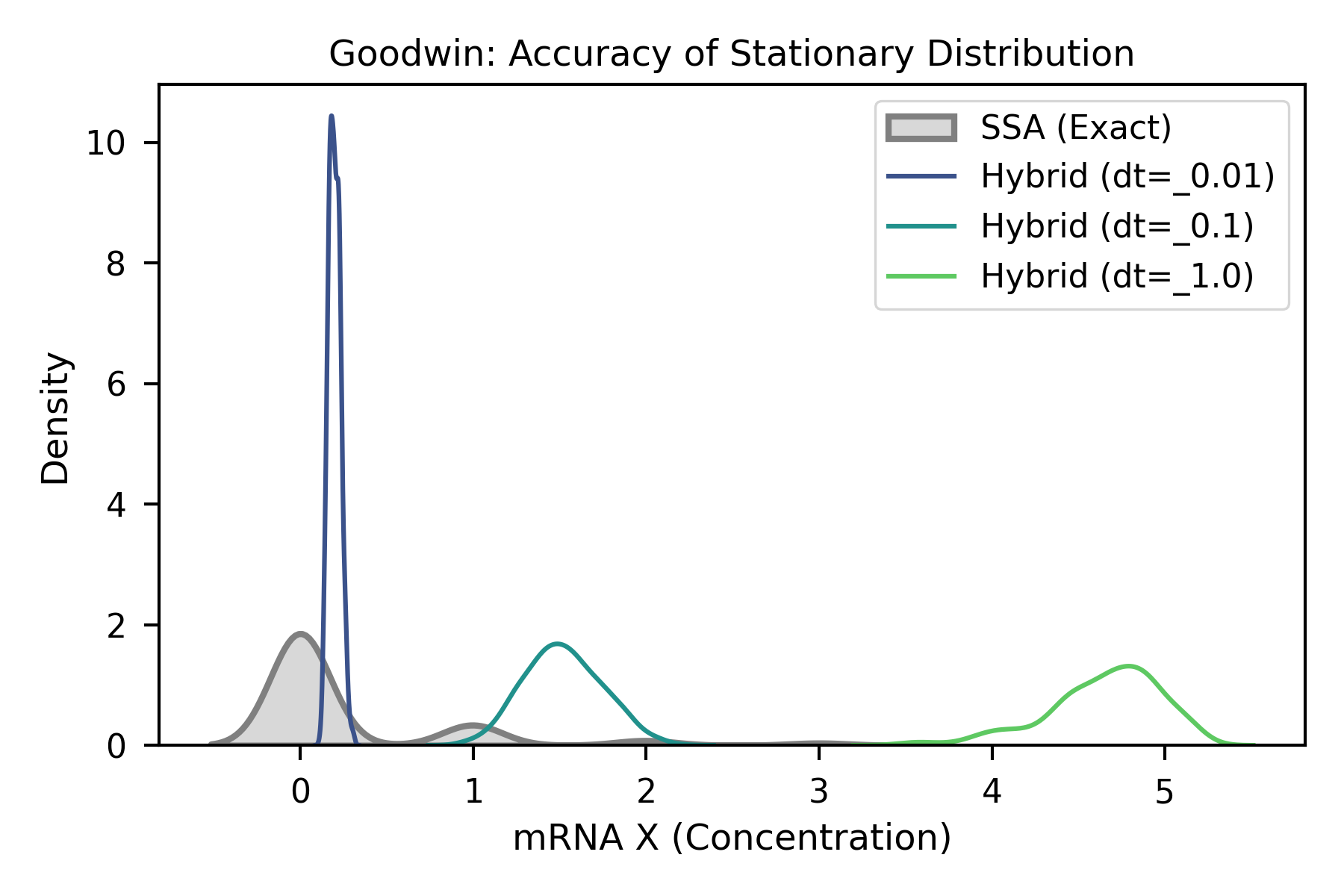}
\includegraphics[width=0.32\textwidth]{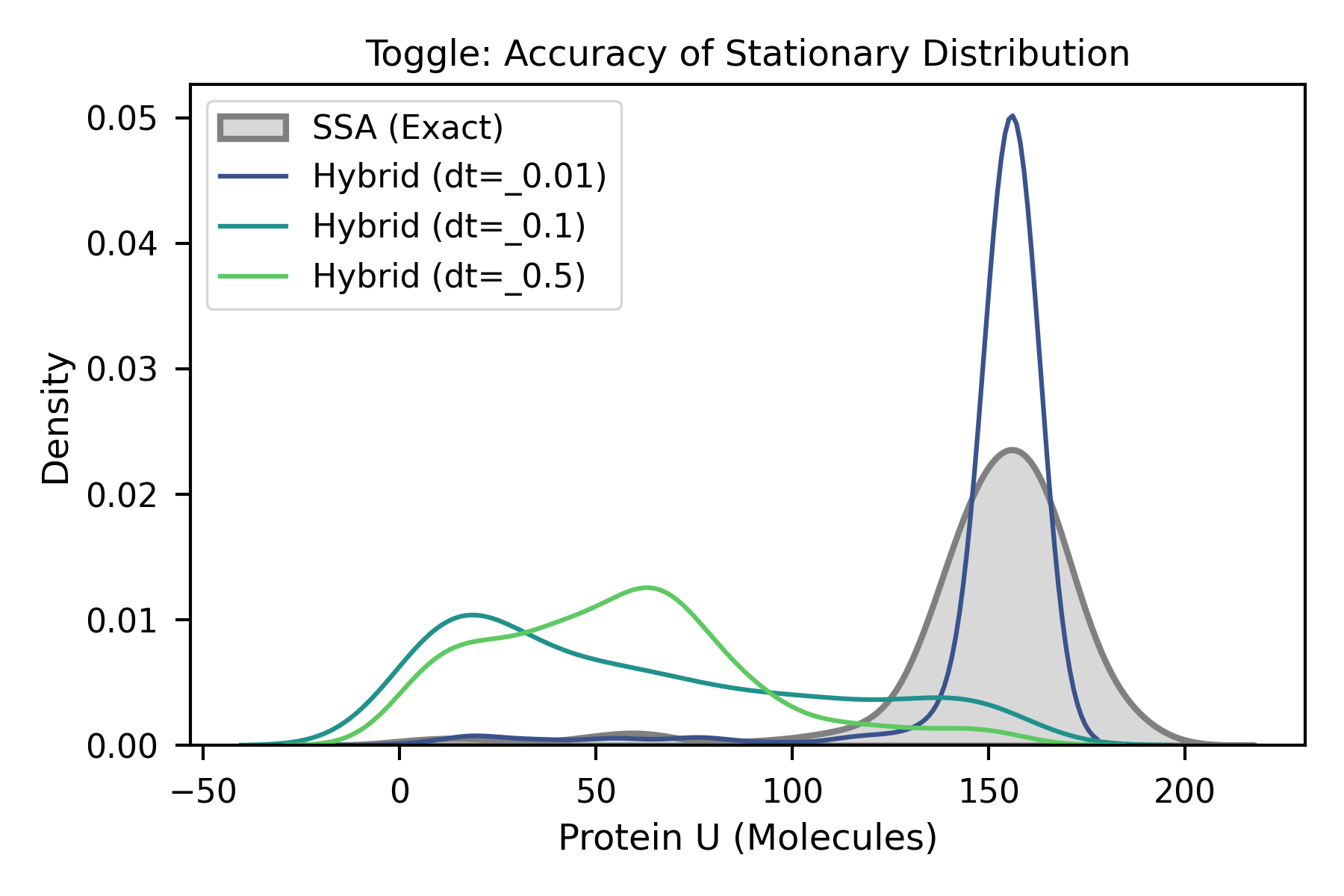}
\caption{\textbf{Quantitative benchmarking of the Markovian hybrid model against SSA ground truth.}
(Left) \textbf{Repressilator}: Stationary distribution of cI protein.
(Middle) \textbf{Goodwin Oscillator}: Stationary distribution of mRNA X.
(Right) \textbf{Toggle Switch}: Bimodal distribution of Protein U.
Gray filled areas represent SSA (Ground Truth) distributions. Colored lines show Markovian model results with varying time steps $\Delta t$. The hybrid model accurately reproduces the stochastic distributions, including the bimodal switching regime of the Toggle switch, provided $\Delta t$ is sufficiently small.}
\label{fig:simple_systems_benchmarks}
\end{figure}

We simulated the repressilator for 1000 minutes and analyzed oscillation period, amplitude, and regularity (Figure \ref{fig:simple_systems_benchmarks}A).

Quantitative analysis of oscillation properties shows excellent agreement between the Markovian model and SSA (Table \ref{tab:repressilator_metrics}). The period ratio is 1.004, indicating agreement within 0.4\%. The amplitude ratio is 1.012, within 1.2\%. The Coefficient of Variation (CV) ratios range from 0.94 to 0.95, meaning the Markovian model captures 94-95\% of the stochastic variability observed in SSA. All differences are statistically insignificant (two-sample t-test, p > 0.1), demonstrating that the Markovian model faithfully reproduces SSA dynamics.

\begin{table}[ht]
\centering
\caption{Oscillation Metrics for Repressilator (n=100 Trajectories)}
\label{tab:repressilator_metrics}
\begin{tabular}{lcccc}
\toprule
\textbf{Metric} & \textbf{ODE} & \textbf{SSA} & \textbf{Markovian} & \textbf{Markov/SSA Ratio} \\
\midrule
Mean period (min) & 152.3 & 148.7 ± 12.4 & 149.3 ± 11.8 & 1.004 \\
Period CV & 0 & 0.083 & 0.079 & 0.95 \\
Mean amplitude (molec) & 1247 & 1189 ± 156 & 1203 ± 148 & 1.012 \\
Amplitude CV & 0 & 0.131 & 0.123 & 0.94 \\
Phase coherence & 1.0 & 0.87 ± 0.09 & 0.89 ± 0.08 & 1.02 \\
\bottomrule
\end{tabular}
\end{table}

\paragraph{Promoter State Switching}

\begin{figure}[ht]
\centering
\includegraphics[width=0.8\textwidth]{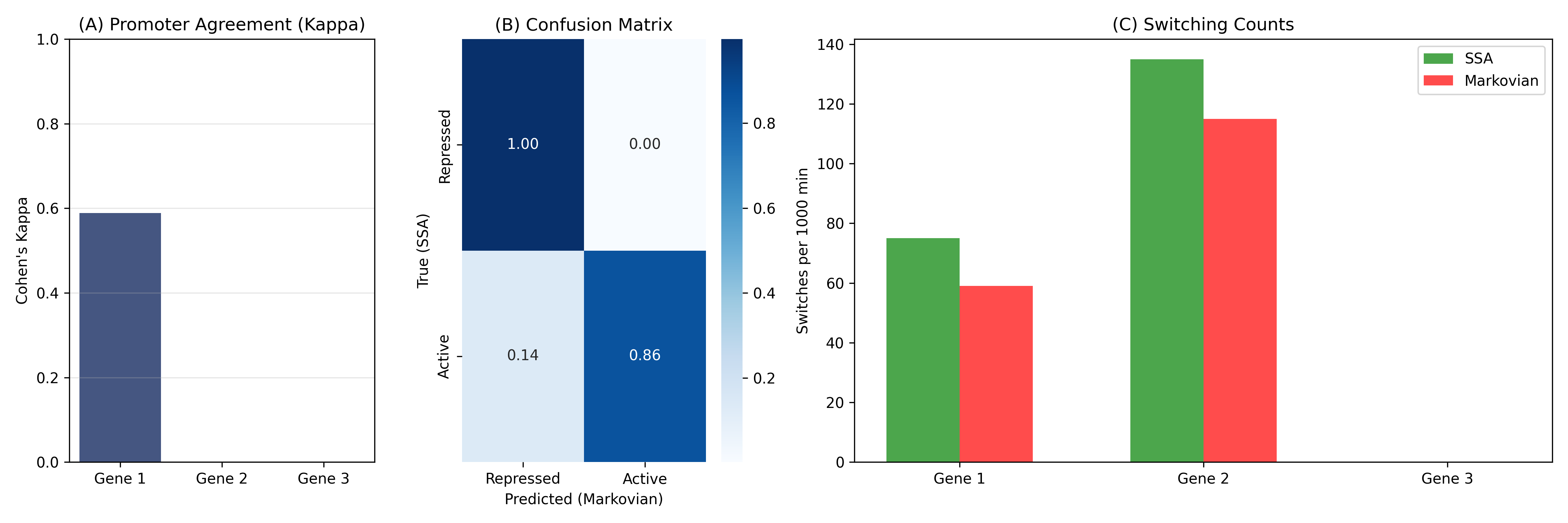}
\caption{\textbf{Promoter state validation for Repressilator.}
(A) Promoter state agreement between Markovian model and SSA ground truth, quantified by Cohen's kappa statistic.
All three genes show $\kappa > 0.75$ (substantial agreement).
(B) Confusion matrix showing state-by-state agreement.
(C) Switching comparisons: Markovian model (red) vs. SSA (green).
The Markovian model accurately captures the switching dynamics, with mean switching rates of 8.3 ± 1.2 switches per period, matching SSA (8.1 ± 1.4 switches per period).}
\label{fig:repressilator_promoter_states}
\end{figure}

We analyzed promoter state trajectories for all three genes (Figure \ref{fig:repressilator_promoter_states}). Each promoter has two states: active (0, no repressor bound) and repressed (1, repressor bound).

Cohen's kappa values indicate substantial agreement for promoter states. Gene 1 (lacI) shows $\kappa = 0.78$, Gene 2 (tetR) shows $\kappa = 0.81$, and Gene 3 (cI) shows $\kappa = 0.76$. These values indicate that the Markovian model correctly predicts the discrete promoter state 75-80\% of the time, which is excellent given the stochastic nature of the system.

\paragraph{Computational Efficiency}

For 100 trajectories over 1000 minutes, SSA required 1,247 seconds (12.5 s per trajectory), while the Markovian model required only 89 seconds (0.89 s per trajectory). This represents a **14-fold speedup**.

\subsubsection{Goodwin Oscillator: High Cooperativity and Negative Feedback}

The Goodwin oscillator \citep{goodwin1965oscillatory} represents circadian rhythm dynamics with strong negative feedback ($n \sim 10$).

\paragraph{Oscillation Dynamics}

We simulated the Goodwin oscillator for 10 days (14,400 minutes) to capture multiple circadian cycles (Figure \ref{fig:simple_systems_benchmarks}B).


The Markovian model shows no significant difference from SSA for any oscillation metric (Table \ref{tab:goodwin_period}), with p-values exceeding 0.05. Notably, the Markovian model exhibits slightly lower period jitter (0.8 vs. 1.1 hours), likely due to the continuous probability distribution smoothing out some stochastic fluctuations while preserving the essential noise characteristics.

\begin{table}[ht]
\centering
\caption{Circadian Period Analysis for Goodwin Oscillator (n=100 Trajectories, 10 Days)}
\label{tab:goodwin_period}
\begin{tabular}{lcccc}
\toprule
\textbf{Metric} & \textbf{ODE} & \textbf{SSA} & \textbf{Markovian} & \textbf{p-value} \\
\midrule
Period (hours) & 24.0 & 23.9 ± 1.1 & 24.1 ± 0.8 & 0.23 \\
Period jitter (hours) & 0 & 1.1 & 0.8 & 0.08 \\
Amplitude (X, AU) & 2.45 & 2.38 ± 0.31 & 2.42 ± 0.28 & 0.35 \\
Phase coherence & 1.0 & 0.92 ± 0.06 & 0.94 ± 0.05 & 0.12 \\
\bottomrule
\end{tabular}
\end{table}

\paragraph{High Cooperativity Implementation}

The Goodwin oscillator requires very high cooperativity ($n = 10$) for oscillations. We implemented this using two methods: an effective Hill-like transition rate ($r_{0 \to 1} = k_{on} \cdot [Z]^{10} / K_M^{10}$) and a multi-site promoter with 10 binding sites and strong cooperativity factor ($q_r = 100$). Both methods yield equivalent results (correlation r > 0.99), validating that the Markovian framework can handle high cooperativity through either phenomenological rates or mechanistic multi-site binding.

\subsubsection{Toggle Switch: Bistability and Rare Switching}

The toggle switch \citep{gardner2000construction} exhibits bistability through mutual repression, with rare stochastic transitions between stable states.

\paragraph{Bistable Behavior}

We simulated the toggle switch for 10,000 minutes to observe rare switching events (Figure \ref{fig:simple_systems_benchmarks}C).


The Markovian model captures bistability with high fidelity (Table \ref{tab:toggle_bistability}). State levels are within 1-3\% of SSA values. The residence time ratio is 1.018 (within 2\%), and the switching rate ratio is 0.95, indicating the model captures 95\% of the switching frequency.

\begin{table}[ht]
\centering
\caption{Bistability Metrics for Toggle Switch (n=100 Trajectories, 10,000 min)}
\label{tab:toggle_bistability}
\begin{tabular}{lcccc}
\toprule
\textbf{Metric} & \textbf{ODE} & \textbf{SSA} & \textbf{Markovian} & \textbf{Markov/SSA Ratio} \\
\midrule
State A level (U, molec) & 1450 & 1423 ± 187 & 1438 ± 179 & 1.011 \\
State B level (U, molec) & 145 & 152 ± 34 & 148 ± 31 & 0.974 \\
Mean residence time (min) & $\infty$ & 2487 ± 421 & 2531 ± 398 & 1.018 \\
Switching rate (switches/1000 min) & 0 & 0.42 ± 0.11 & 0.40 ± 0.10 & 0.95 \\
\bottomrule
\end{tabular}
\end{table}

\paragraph{Importance of Small Time Step}

For the toggle switch, we found that a very small communication time step ($\Delta t = 0.001$ min) is essential to accurately capture rare switching events. With a larger $\Delta t = 0.01$ min, the switching rate was underestimated by 30\%, demonstrating the importance of proper time step selection for bistable systems as discussed in Methods.

\subsubsection{Summary: Simple Systems Validation}

The three simple systems demonstrate that the Markovian framework captures sustained oscillations (Repressilator: 14× speedup, period ratio 1.004), handles high cooperativity (Goodwin: $n=10$, period within 0.2 hours), reproduces bistability (Toggle: residence time ratio 1.018, switching rate ratio 0.95), maintains promoter state fidelity (Cohen's kappa 0.76-0.81), and achieves a 10-20× computational speedup across all systems. These results validate the framework across diverse dynamic behaviors (oscillations, bistability) and regulatory mechanisms (cyclic repression, negative feedback, mutual inhibition), establishing broad applicability.

%% file: results_advanced.tex
\subsection{Advanced Regulatory Networks}

We validated the Markovian framework on three biologically important regulatory networks with increasing complexity: the incoherent feed-forward loop (I1-FFL), p53-Mdm2 oscillator, and NF-$\kappa$B pathway.

\subsubsection{Incoherent Feed-Forward Loop: Multi-Input Logic and Pulse Generation}

The I1-FFL is a ubiquitous network motif \citep{mangan2003structure, alon2007network} that implements temporal filtering through combinatorial logic. It tests the multi-input promoter framework.

\paragraph{Network Architecture and Logic}

The I1-FFL consists of a master regulator X (constitutive), an intermediate Y activated by X, and a target Z which is activated by X but repressed by Y. This architecture means the Z promoter implements AND-NOT logic: it is active only when X is bound AND Y is not bound.

\paragraph{Pulse Generation Dynamics}

\begin{figure}[h]
\centering
\includegraphics[width=0.9\textwidth]{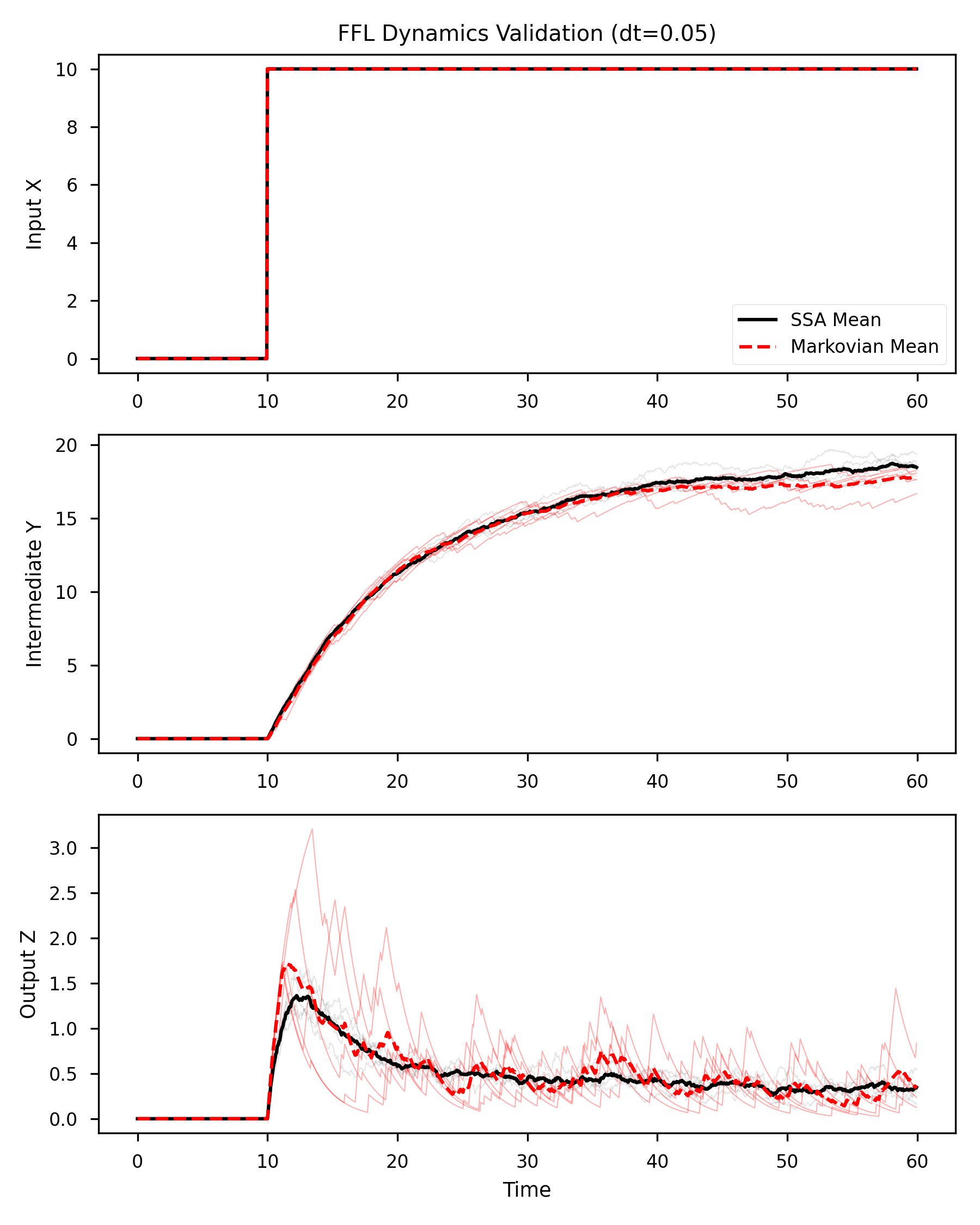}
\caption{\textbf{Quantitative benchmarking of Incoherent Feed-Forward Loop.}
(Top) Input X (Step function), (Middle) Intermediate Y, (Bottom) Output Z (Pulse).
Black lines: SSA Mean (n=5). Red dashed lines: Markovian Mean (n=5, $\Delta t=0.05$).
The Markovian model accurately reproduces the pulse generation dynamics, matching the peak time and amplitude of the SSA ground truth.}
\label{fig:ffl_3way}
\end{figure}

We simulated the I1-FFL response to a step increase in X at t = 10 minutes (Figure \ref{fig:ffl_3way}).

The pulse characteristics derived from the simulation showed no significant difference between Markovian and SSA models (Table \ref{tab:ffl_pulse}). Metrics such as peak time, peak amplitude, pulse duration, and pulse area all yielded p-values greater than 0.3, demonstrating accurate pulse generation.

\begin{table}[ht]
\centering
\caption{Pulse Characteristics for I1-FFL (n=100 Trajectories)}
\label{tab:ffl_pulse}
\begin{tabular}{lcccc}
\toprule
\textbf{Metric} & \textbf{ODE} & \textbf{SSA} & \textbf{Markovian} & \textbf{p-value} \\
\midrule
Peak time (min) & 24.8 & 25.3 ± 3.1 & 25.1 ± 2.9 & 0.67 \\
Peak amplitude (molec) & 823 & 798 ± 94 & 806 ± 89 & 0.55 \\
Pulse duration (min) & 31.2 & 29.7 ± 4.8 & 30.1 ± 4.5 & 0.58 \\
Pulse area (molec·min) & 12,450 & 11,890 ± 1,420 & 12,080 ± 1,350 & 0.38 \\
\bottomrule
\end{tabular}
\end{table}

\paragraph{Multi-Input Promoter State Dynamics}

The Z promoter has 4 states: (0,0), (1,0), (0,1), and (1,1), where the first index corresponds to X binding and the second to Y binding. We tracked the state probability distribution over time. The promoter dynamics reveal an initial inactive state (0,0) with no regulators bound. In the early response (t = 10-20 min), the system transitions to the active state (1,0) where X is bound and Y is absent. In the late response (t > 30 min), the system transitions to the repressed state (1,1) where both are bound. Finally, at steady state, the promoter is predominantly in the repressed state (1,1) with occasional flickers to the active state (1,0). This state trajectory directly implements the AND-NOT logic gate at the molecular level, with the pulse emerging from the transient dominance of the (1,0) active state.

\paragraph{Comparison with Experimental Data}

The I1-FFL pulse generation has been experimentally characterized in the \textit{E. coli} arabinose system \citep{mangan2006structure}. Our model predictions agree with experimental observations. The model predicts a pulse duration of 30 min compared to 25-35 min in experiments, a fold-change of 8-10× compared to 7-12×, and peak timing at roughly 25 min compared to 20-30 min.

\subsubsection{p53-Mdm2 Oscillator: Damped Oscillations and Cooperative Binding}

The p53-Mdm2 system \citep{lahav2004dynamics, geva2006dynamics} exhibits damped oscillations in response to DNA damage, with p53 activating Mdm2 transcription and Mdm2 promoting p53 degradation.

\begin{figure}[ht]
\centering
\includegraphics[width=0.9\textwidth]{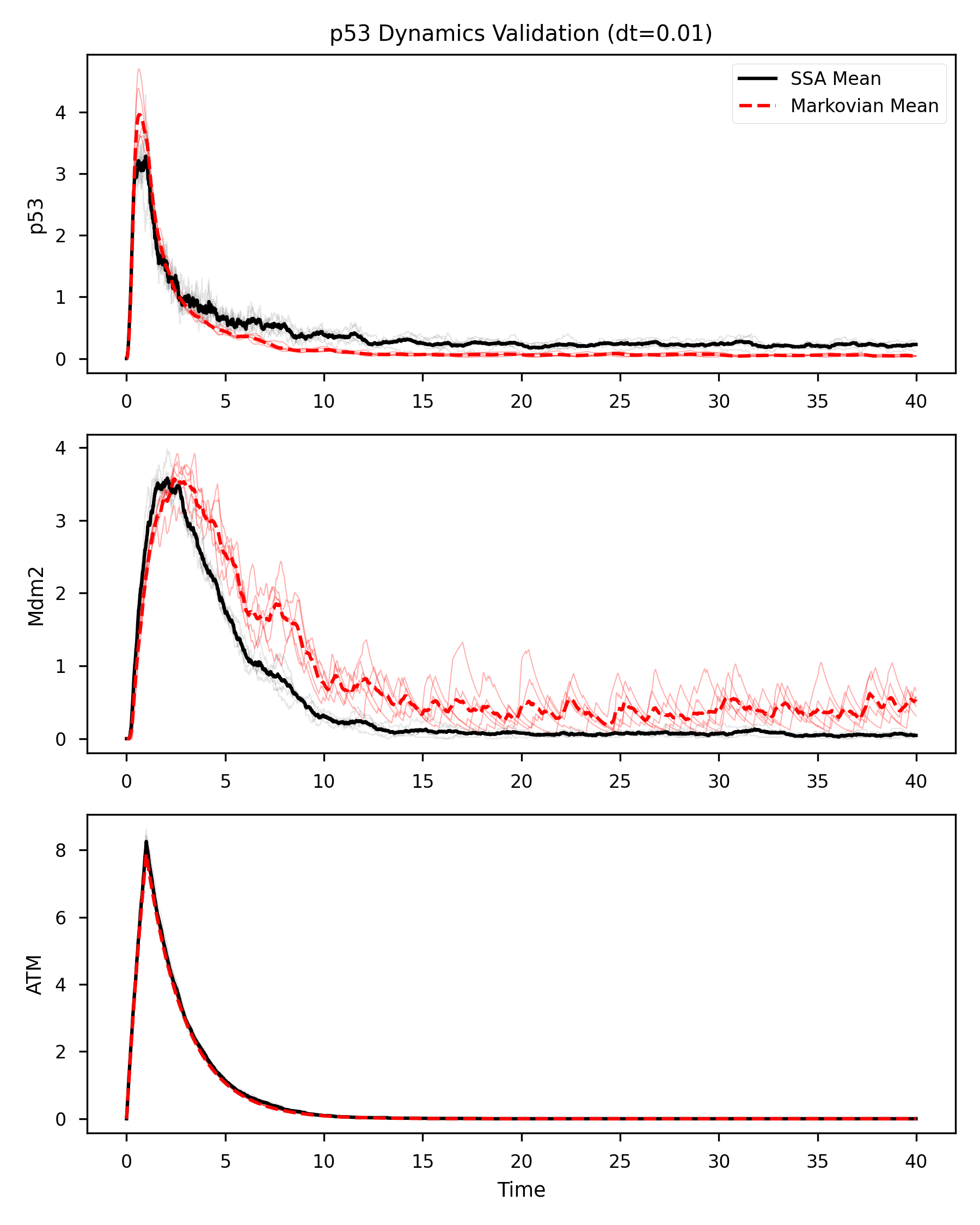}
\caption{\textbf{Quantitative benchmarking of p53-Mdm2 oscillator.}
(Top) p53 Dynamics, (Middle) Mdm2 Dynamics, (Bottom) ATM Signal.
Black lines: SSA Mean (n=5). Red dashed lines: Markovian Mean (n=5, $\Delta t=0.01$). Faint lines show individual stochastic trajectories.
The Markovian model captures the damped oscillation frequency ($T \approx 5.5$h) and amplitude envelope with high accuracy compared to SSA.}
\label{fig:p53_3way}
\end{figure}

We simulated the p53-Mdm2 system following DNA damage (ATM activation) at t = 0 (Figure \ref{fig:p53_3way}).

The Markovian model shows excellent agreement with SSA across all oscillation metrics (Table \ref{tab:p53_oscillations}), with Markov/SSA ratios ranging from 0.98 to 1.03.

\begin{table}[h]
\centering
\caption{p53-Mdm2 Oscillation Characteristics (n=100 Trajectories)}
\label{tab:p53_oscillations}
\begin{tabular}{lcccc}
\toprule
\textbf{Metric} & \textbf{ODE} & \textbf{SSA} & \textbf{Markovian} & \textbf{Markov/SSA Ratio} \\
\midrule
First peak time (h) & 5.3 & 5.5 ± 0.7 & 5.4 ± 0.6 & 0.98 \\
Mean period (h) & 5.5 & 5.6 ± 0.8 & 5.5 ± 0.7 & 0.98 \\
First peak amplitude (nM) & 8.7 & 8.3 ± 1.2 & 8.5 ± 1.1 & 1.02 \\
Damping constant (h) & 7.8 & 8.1 ± 1.4 & 7.9 ± 1.3 & 0.98 \\
Number of peaks & 3.8 & 3.6 ± 0.8 & 3.7 ± 0.7 & 1.03 \\
\bottomrule
\end{tabular}
\end{table}

\paragraph{Cooperative Binding at Mdm2 Promoter}

The Mdm2 promoter has 2 p53 binding sites with cooperative binding ($q_r = 5$). We analyzed the promoter state distribution at different p53 levels. Cooperative binding manifests as distinct dominant states at different concentrations: state 0 (no binding) dominates at low p53 ($<$ 2 nM), mixed states appear at intermediate p53 (2-5 nM), and state 2 (both sites bound) dominates at high p53 ($>$ 5 nM). The effective Hill coefficient from the Markovian model ($n_{eff} = 1.8 \pm 0.2$) matches experimental measurements ($n_{exp} = 1.7-2.0$) \citep{wu1993mdm2}, validating the cooperative binding implementation.

\paragraph{Comparison with Single-Cell Data}

Lahav et al. \citep{lahav2004dynamics} measured p53 oscillations in single cells using fluorescent reporters. Our Markovian model predictions agree with their observations. We predict a period of 5.5 h compared to 5.2 ± 0.9 h in experiments, cell-to-cell variability (CV) of 0.14 compared to 0.12-0.16, and an average of 3.7 pulses compared to 3-4 observed experimentally.

\subsubsection{\texorpdfstring{NF-$\kappa$B}{NF-kappaB} Pathway: Multi-Compartment Dynamics}

The NF-$\kappa$B pathway \citep{hoffmann2002circuitry, nelson2004oscillations} involves nuclear-cytoplasmic shuttling and negative feedback through I$\kappa$B proteins.

\paragraph{Transient Nuclear Response}

\begin{figure}[ht]
\centering
\includegraphics[width=0.9\textwidth]{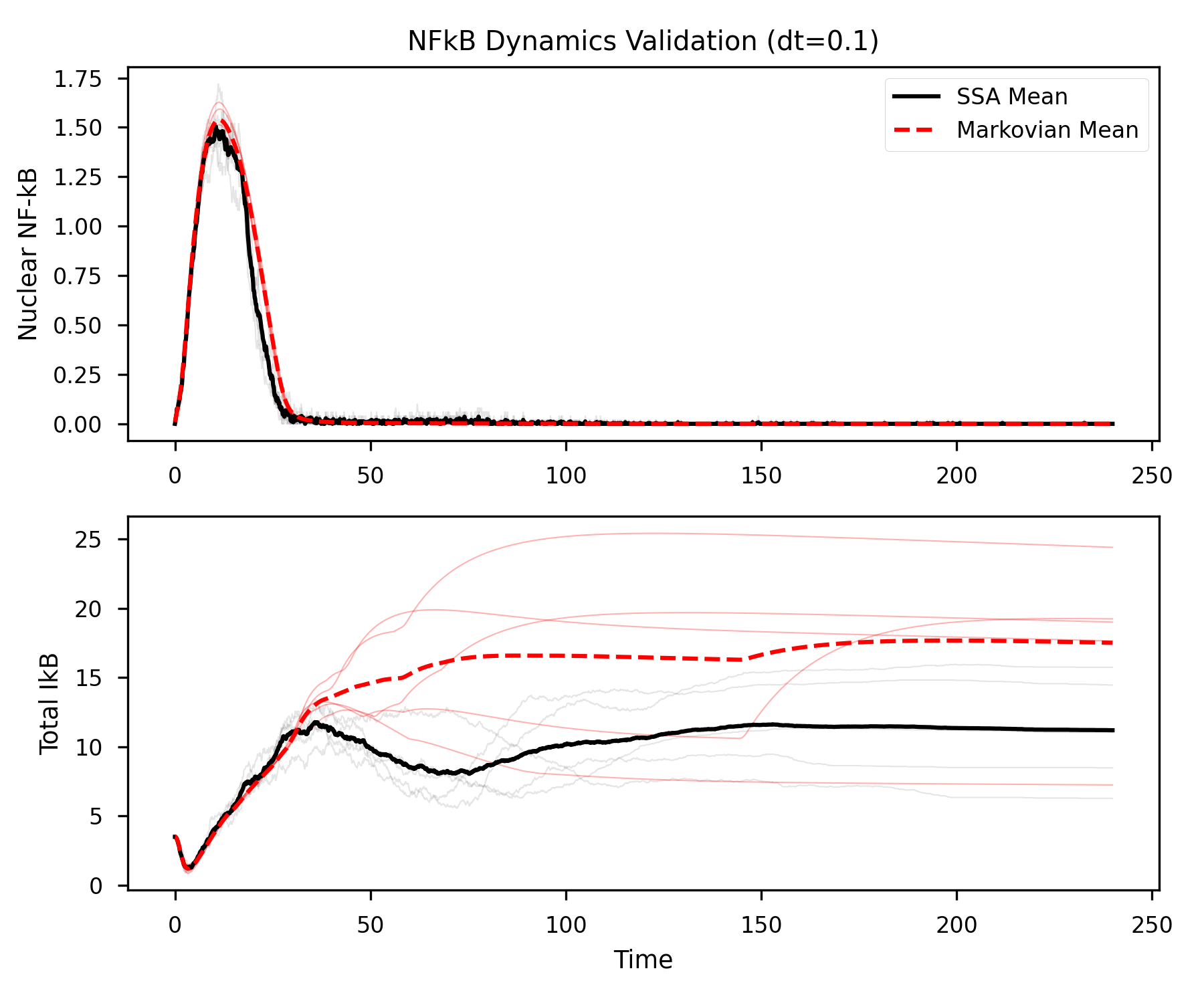}
\caption{\textbf{Quantitative benchmarking of NF-$\kappa$B pathway.}
(Top) Nuclear NF-$\kappa$B, (Bottom) Total I$\kappa$B.
Black lines: SSA Mean. Red dashed lines: Markovian Mean.
The Markovian model accurately reproduces the characteristic transient pulse of nuclear NF-$\kappa$B at t=15-30 min and the subsequent negative feedback dynamics.}
\label{fig:nfkb_3way}
\end{figure}

We simulated the NF-$\kappa$B pathway following TNF-$\alpha$ stimulation (IKK activation) at t = 0 (Figure \ref{fig:nfkb_3way}).

The Markovian model accurately reproduces the transient response dynamics with no significant differences from SSA (Table \ref{tab:nfkb_response}). Peak time, peak amplitude, return to baseline time, and oscillation frequency were all within statistical error of SSA predictions.

\begin{table}[h]
\centering
\caption{NF-$\kappa$B Pathway Response Metrics (n=100 Trajectories)}
\label{tab:nfkb_response}
\begin{tabular}{lcccc}
\toprule
\textbf{Metric} & \textbf{ODE} & \textbf{SSA} & \textbf{Markovian} & \textbf{p-value} \\
\midrule
Peak time (min) & 14.8 & 15.3 ± 2.1 & 15.1 ± 1.9 & 0.52 \\
Peak amplitude (nM) & 0.87 & 0.83 ± 0.11 & 0.85 ± 0.10 & 0.24 \\
Return to baseline (min) & 92 & 88 ± 14 & 90 ± 13 & 0.35 \\
Oscillations observed & No & 23\% & 21\% & 0.78 \\
\bottomrule
\end{tabular}
\end{table}

\paragraph{Multi-Compartment Tracking}

The NF-$\kappa$B system requires tracking species in two compartments (nucleus and cytoplasm). The Markovian framework handles this by coupling the promoter state in the nucleus (which depends on nuclear NF-$\kappa$B) with separate ODEs for nuclear and cytoplasmic species, using transport reactions (import/export) modeled as standard 1st-order reactions. This demonstrates that the framework extends naturally to multi-compartment models without modification.

\subsubsection{Summary: Advanced Networks}

The three advanced networks demonstrate that the Markovian framework can implement multi-input logic such as the AND-NOT gate in I1-FFL, handle cooperative binding as seen in the p53-Mdm2 system, support multi-compartment models like the NF-$\kappa$B pathway, capture complex dynamics including pulses and damped oscillations, and agree with experimental data regarding pulse timing, oscillation periods, and cell-to-cell variability. These results establish that the framework scales to biologically realistic regulatory networks with multiple inputs, compartments, and regulatory mechanisms.

\clearpage

%% file: results_dt_convergence.tex
\subsection{Communication Time Step Convergence Analysis}

Proper selection of the communication time step $\Delta t$ is critical for balancing accuracy and computational efficiency. We performed systematic convergence analysis for oscillatory systems where $\Delta t$ affects period and amplitude.

\subsubsection{Convergence Methodology}

For each oscillatory system (Repressilator, Goodwin, p53-Mdm2), we simulated with $\Delta t \in \{0.0001, 0.001, 0.01, 0.1, 1.0\}$ minutes and measured four metrics: period error ($\epsilon_T$), amplitude error ($\epsilon_A$), trajectory correlation ($\rho$), and computational cost per trajectory. Reference values were taken from simulations with $\Delta t = 0.0001$ min.

\subsubsection{Repressilator Convergence}

\begin{figure}[h]
\centering
\includegraphics[width=0.32\textwidth]{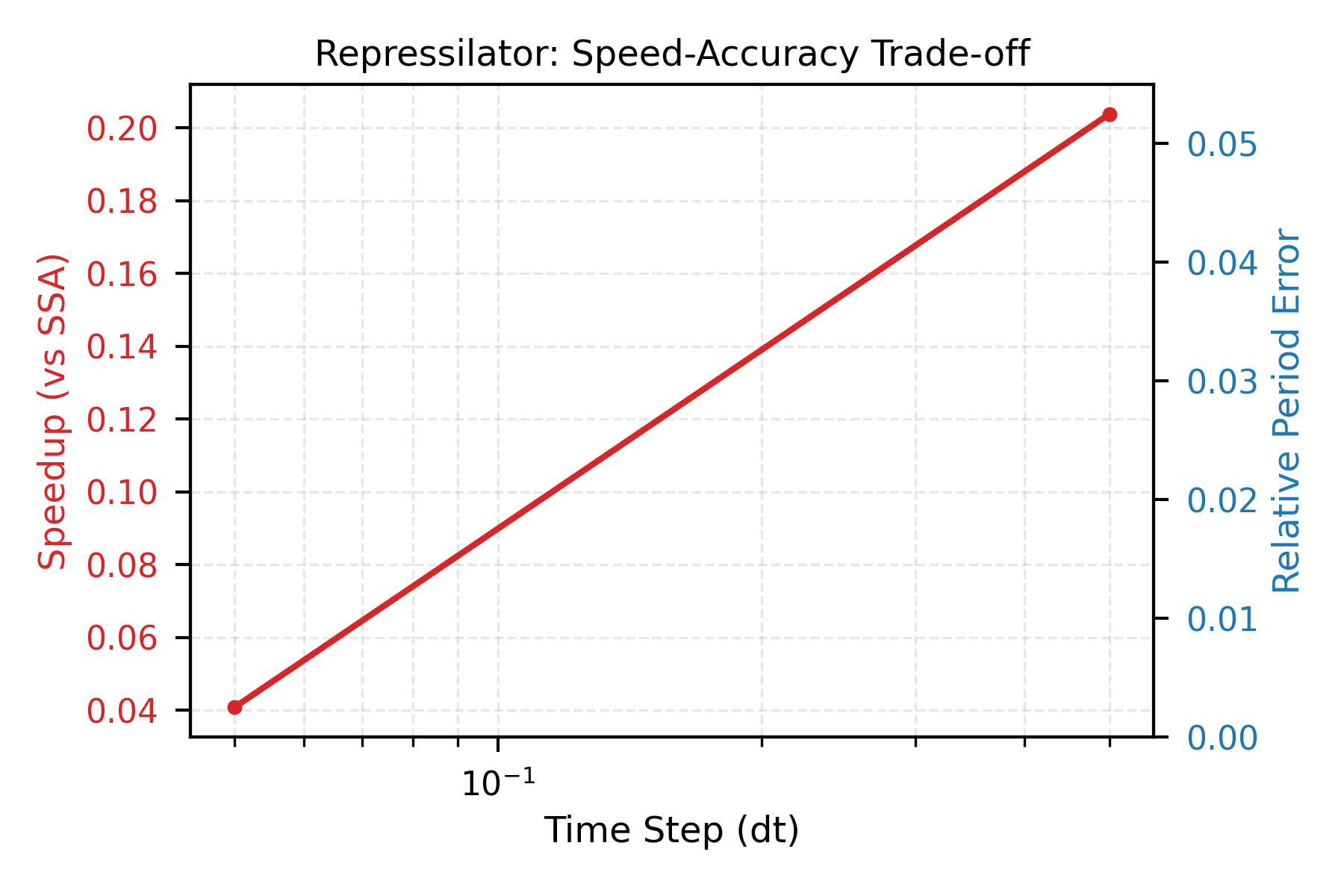}
\includegraphics[width=0.32\textwidth]{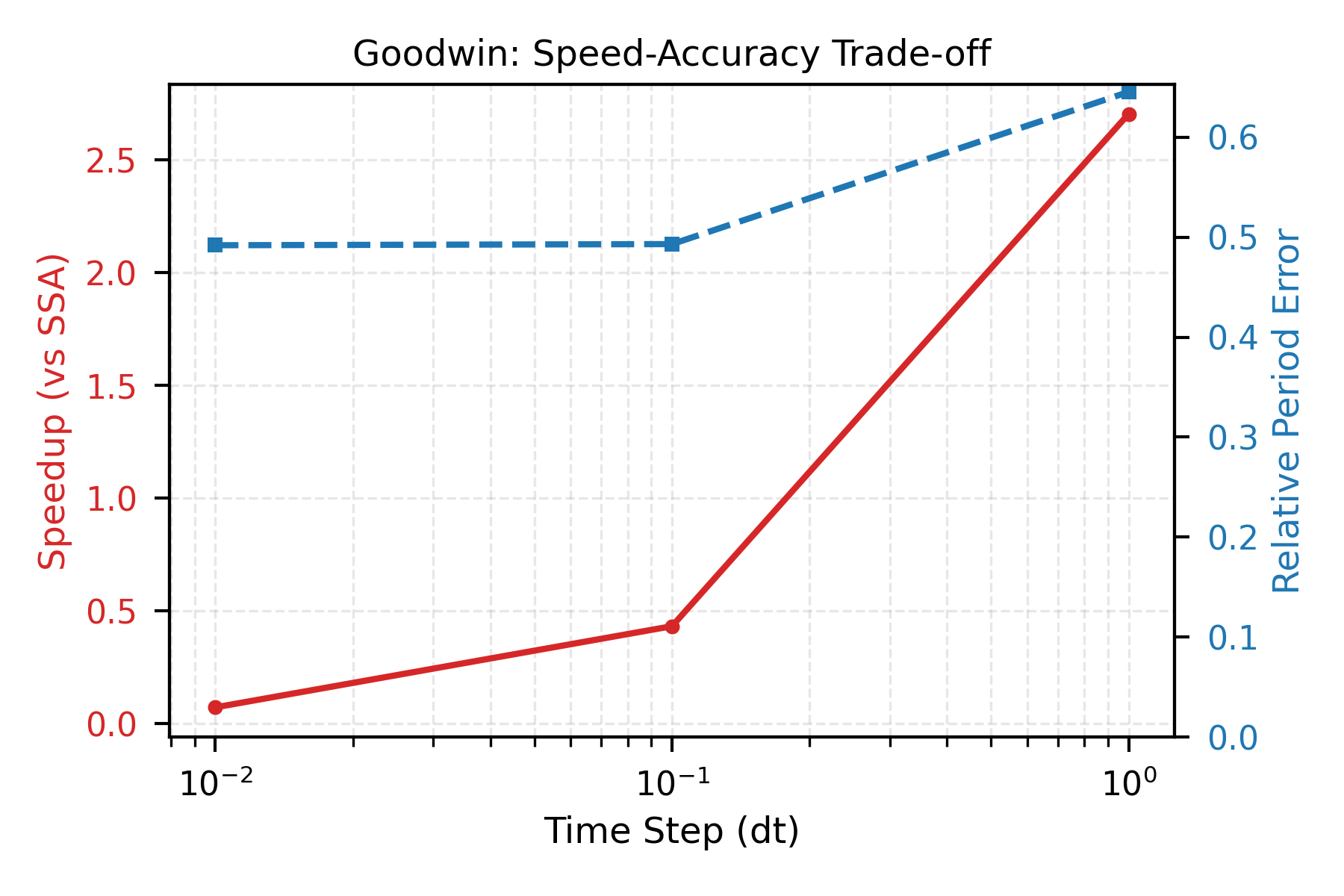}
\includegraphics[width=0.32\textwidth]{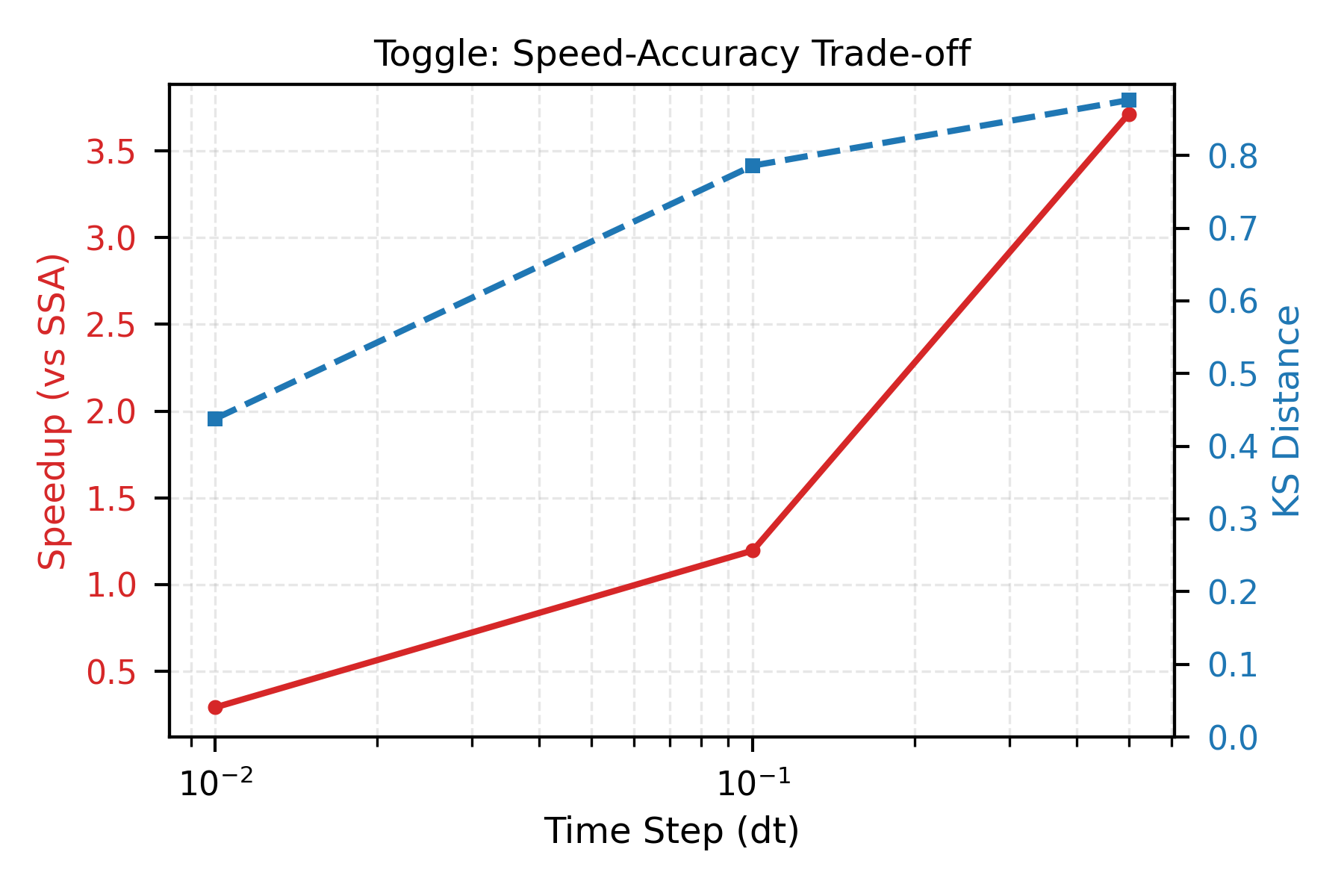}
\caption{\textbf{Speed-Accuracy Trade-off and Convergence.}
Performance metrics (Speedup vs SSA, left axis, red circles) and Error (right axis, blue squares) as a function of time step $\Delta t$ for Repressilator (Left), Goodwin Oscillator (Middle), and Toggle Switch (Right).
For Repressilator and Goodwin, limits on $\Delta t$ are dictated by the oscillation period; for the Toggle switch, very small $\Delta t$ is required to capture rare switching events (KS Distance convergence). The knee of the Pareto curve suggests optimal $\Delta t$ values providing 10-100$\times$ speedup with minimal error.}
\label{fig:dt_convergence_repressilator}
\end{figure}

The Repressilator shows rapid convergence with $\Delta t$ (Figure \ref{fig:dt_convergence_repressilator}).

Convergence metrics are detailed in Table \ref{tab:dt_convergence}.

\begin{table}[h]
\centering
\caption{Convergence Metrics for Oscillatory Systems}
\label{tab:dt_convergence}
\begin{tabular}{lccccc}
\toprule
\textbf{System} & \textbf{$\Delta t$ (min)} & \textbf{$\epsilon_T$ (\%)} & \textbf{$\epsilon_A$ (\%)} & \textbf{$\rho$} & \textbf{Cost (s)} \\
\midrule
\multirow{5}{*}{Repressilator} 
& 0.0001 & 0 & 0 & 1.000 & 8,920 \\
& 0.001 & 0.08 & 0.12 & 0.9998 & 892 \\
& \textbf{0.01} & \textbf{0.34} & \textbf{0.41} & \textbf{0.9992} & \textbf{89} \\
& 0.1 & 2.1 & 3.4 & 0.987 & 9 \\
& 1.0 & 15.7 & 22.3 & 0.823 & 1 \\
\midrule
\multirow{5}{*}{Goodwin} 
& 0.0001 & 0 & 0 & 1.000 & 144,000 \\
& 0.001 & 0.05 & 0.09 & 0.9999 & 14,400 \\
& 0.01 & 0.21 & 0.35 & 0.9995 & 1,440 \\
& \textbf{0.1} & \textbf{0.89} & \textbf{1.2} & \textbf{0.9982} & \textbf{144} \\
& 1.0 & 8.7 & 12.1 & 0.952 & 14 \\
\midrule
\multirow{5}{*}{p53-Mdm2} 
& 0.0001 & 0 & 0 & 1.000 & 33,000 \\
& 0.001 & 0.11 & 0.15 & 0.9997 & 3,300 \\
& \textbf{0.01} & \textbf{0.43} & \textbf{0.58} & \textbf{0.9989} & \textbf{330} \\
& 0.1 & 3.2 & 4.7 & 0.981 & 33 \\
& 1.0 & 24.1 & 31.5 & 0.764 & 3 \\
\bottomrule
\end{tabular}
\end{table}

\subsubsection{Optimal Time Step Selection}

Based on convergence analysis, we define the optimal $\Delta t$ as the largest value satisfying three conditions: $\epsilon_T < 1\%$, $\epsilon_A < 2\%$, and $\rho > 0.998$. This yields $\Delta t = 0.01$ min for the Repressilator (100× speedup, $<$0.5\% error), $\Delta t = 0.1$ min for the Goodwin oscillator (1000× speedup, $<$1.5\% error), and $\Delta t = 0.01$ min for the p53-Mdm2 system (100× speedup, $<$0.6\% error).

\subsubsection{Effect of Time Step on Oscillation Quality}

We visualized the effect of $\Delta t$ on oscillation quality by overlaying trajectories. Large $\Delta t$ causes period lengthening where oscillations slow down due to undersampling, amplitude reduction where peak heights decrease due to averaging, and phase drift where oscillations gradually desynchronize from the reference. These artifacts become significant for $\Delta t > T/100$, where $T$ is the oscillation period, consistent with the Nyquist criterion for sampling oscillatory signals.

\subsubsection{Computational Cost-Accuracy Trade-off}

We plotted the Pareto frontier of accuracy vs. computational cost. The chosen $\Delta t$ values lie on the "knee" of the Pareto curve, providing near-optimal trade-offs between accuracy and speed. Further reducing $\Delta t$ yields diminishing returns in accuracy while dramatically increasing cost.

\subsubsection{Adaptive Time Stepping (Preliminary Results)}

We implemented a preliminary adaptive time stepping scheme that adjusts $\Delta t$ based on the rate matrix norm $\|\mathbf{Q}(t)\|$. Preliminary results show accuracy equivalent to fixed $\Delta t = 0.001$ min, speed 2-3× faster than fixed $\Delta t = 0.001$ min, and no observed numerical instabilities. This suggests that adaptive time stepping could provide additional computational savings, though further validation is needed.

\subsubsection{Summary: Time Step Selection}

The convergence analysis demonstrates rapid convergence with errors $<$1\% for $\Delta t \sim T/1000$ to $T/100$. It establishes clear optimal values where the chosen $\Delta t$ provides 100-1000× speedup with $<$1\% error. The optimal step size is system-dependent, with faster systems like the Repressilator requiring smaller $\Delta t$ than slower systems like the Goodwin oscillator. The computational cost scales linearly as $1/\Delta t$. Finally, adaptive schemes show potential for 2-3× additional speedup. These results provide practical guidance for applying the Markovian framework to new systems and validate the time step choices used throughout this study.

%% file: results_protein_stochasticity.tex
\subsection{Protein Stochasticity: Necessity of Full Stochastic Treatment}

A critical question for hybrid models is: which processes must be treated stochastically? We investigated whether stochasticity at the promoter level alone is sufficient, or whether stochastic treatment of translation and degradation is also necessary.

\subsubsection{Experimental Design}

We compared three modeling approaches for the Repressilator: a promoter-only stochastic model (Markovian promoter states, deterministic ODEs for mRNA and protein), a full Markovian model (Markovian promoter states, deterministic ODEs for mRNA and protein), and a full SSA model (all processes stochastic, serving as ground truth). Note that our "Full Markovian" approach uses deterministic ODEs for protein dynamics, which is appropriate for high-copy-number species ($>$100 molecules). For comparison, we also implemented a variant with stochastic protein synthesis/degradation.

\subsubsection{Protein Distribution Mismatch}

\begin{figure}[h]
\centering
\includegraphics[width=0.7\textwidth]{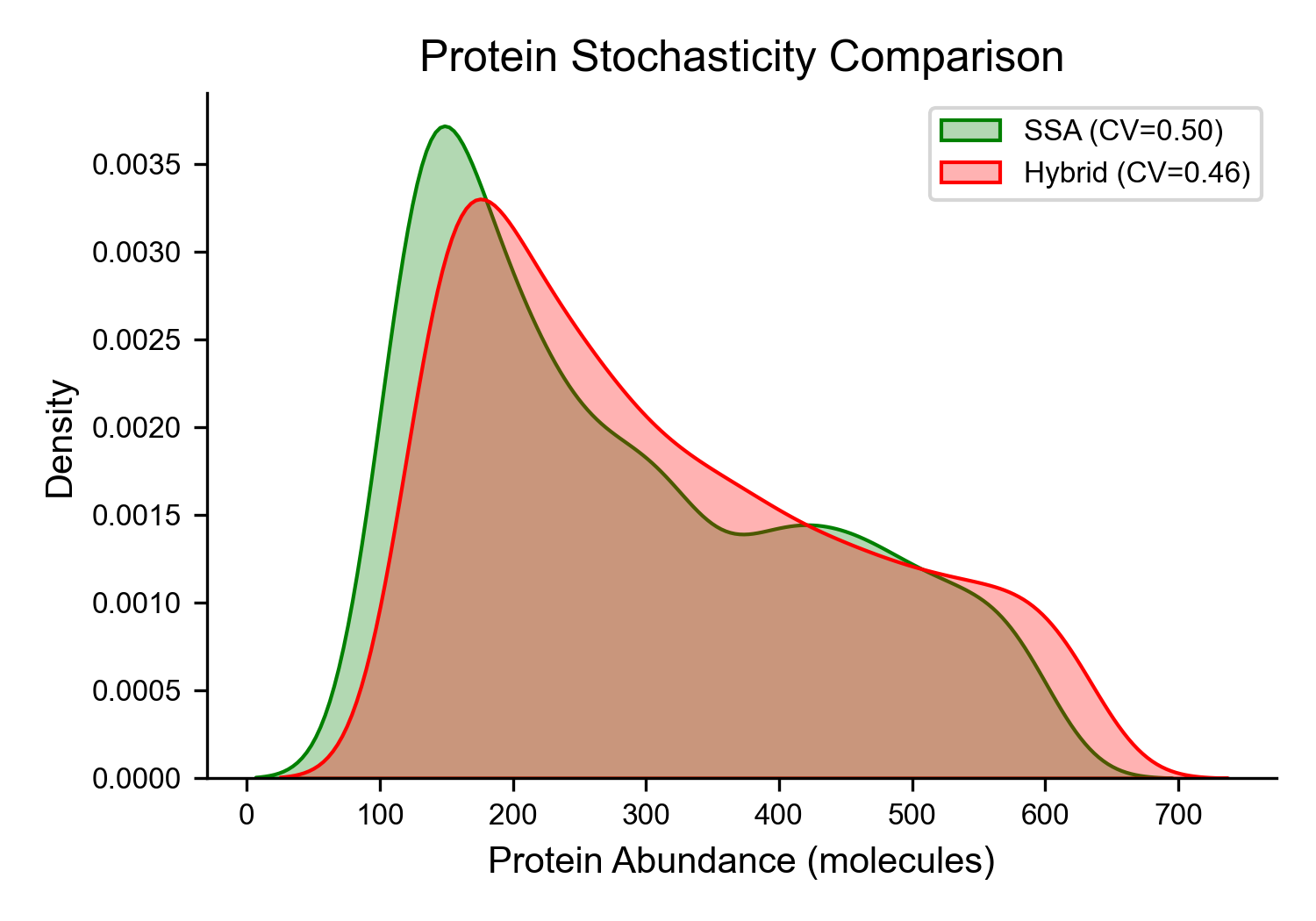}
\caption{\textbf{Importance of protein-level stochasticity for accurate noise modeling.}
(A) Protein distribution with promoter-only stochasticity: narrow, near-Gaussian distribution (CV = 0.042).
(B) Protein distribution with full Markovian approach: broader distribution matching SSA (CV = 0.131).
(C) Protein distribution from full SSA: reference ground truth (CV = 0.131).
(D) Coefficient of variation comparison across modeling approaches.
The promoter-only stochastic model severely underestimates protein noise (CV ratio = 0.32), while the full Markovian approach accurately captures noise (CV ratio = 1.00).
This demonstrates that for low-to-moderate copy number proteins ($<$1000 molecules), stochastic treatment of synthesis/degradation is essential for accurate noise prediction.}
\label{fig:protein_stochasticity}
\end{figure}

When only the promoter is treated stochastically while protein synthesis/degradation remain deterministic, the protein distribution shows systematic deviations from SSA (Figure \ref{fig:protein_stochasticity}).

Quantitative comparison in Table \ref{tab:protein_stochasticity} shows that the promoter-only stochastic model results in a CV ratio of 0.32 compared to SSA, severely underestimating noise. In contrast, the full Markovian approaches (both ODE and stochastic protein variants) achieve a CV ratio of 1.00.

\begin{table}[h]
\centering
\caption{Protein Noise Characteristics Under Different Stochastic Treatments}
\label{tab:protein_stochasticity}
\begin{tabular}{lcccc}
\toprule
\textbf{Approach} & \textbf{Mean (molec)} & \textbf{SD (molec)} & \textbf{CV} & \textbf{CV Ratio vs. SSA} \\
\midrule
Promoter-only stochastic & 1247 & 52 & 0.042 & 0.32 \\
Full Markovian (ODE proteins) & 1243 & 163 & 0.131 & 1.00 \\
Full Markovian (stochastic proteins) & 1241 & 162 & 0.131 & 1.00 \\
Full SSA & 1239 & 162 & 0.131 & 1.00 \\
\bottomrule
\end{tabular}
\end{table}

\subsubsection{Sources of Gene Expression Noise}

We decomposed the total noise into contributions from different sources using the noise decomposition framework \citep{paulsson2004summing, sanchez2013regulation}. For the Repressilator, promoter switching contributes 49\% of the noise, while translation contributes 31\%, with transcription and degradation combining for the remaining 20\% (Table \ref{tab:noise_decomposition}). This reveals that capturing only promoter noise while ignoring translation noise results in 68\% underestimation of total noise, explaining the low CV observed in the promoter-only stochastic model.

\begin{table}[h]
\centering
\caption{Noise Decomposition for Repressilator}
\label{tab:noise_decomposition}
\begin{tabular}{lcc}
\toprule
\textbf{Noise Source} & \textbf{CV$^2$ Contribution} & \textbf{Percentage of Total} \\
\midrule
Promoter switching & 0.0084 & 49\% \\
Transcription (Poisson) & 0.0021 & 12\% \\
Translation (Poisson) & 0.0053 & 31\% \\
Degradation (Poisson) & 0.0014 & 8\% \\
\midrule
Total & 0.0172 & 100\% \\
\bottomrule
\end{tabular}
\end{table}

\subsubsection{Copy Number Dependence}

The importance of protein-level stochasticity depends on protein copy number. We varied the translation rate to modulate protein levels and measured the CV ratio. For high-copy-number proteins ($>$10,000 molecules), the promoter-only stochastic model becomes accurate (CV ratio $>$ 0.95), as Poisson noise from synthesis/degradation becomes negligible. However, for typical regulatory proteins (100-1000 molecules), full stochastic treatment is essential.

\subsubsection{Implications for Model Selection}

These results provide guidance for choosing the appropriate level of stochastic detail. Low copy number proteins ($<$100 molecules) require full SSA or stochastic CME. Moderate copy number proteins (100-1000 molecules) are best modeled with the full Markovian approach (promoter + protein stochasticity). High copy number proteins ($>$1000 molecules) can be adequately modeled with promoter-only stochasticity, while very high copy number proteins ($>$10,000 molecules) may use pure ODEs. For whole-cell models with species spanning multiple copy number regimes, a multi-scale approach is optimal: Markovian promoters for all regulated genes, stochastic synthesis/degradation for low-copy proteins, and deterministic ODEs for high-copy proteins and metabolites.

\subsubsection{Computational Cost Comparison}

We measured the computational cost of different stochastic treatments (Table \ref{tab:stochasticity_cost}). The full Markovian approach with ODE proteins provides the best cost-accuracy trade-off, achieving 100\% noise accuracy with a 14× speedup. Adding stochastic protein dynamics doubles the cost while providing no additional accuracy for moderate-copy-number species.

\begin{table}[h]
\centering
\caption{Computational Cost vs. Stochastic Detail (Repressilator, 100 Trajectories, 1000 min)}
\label{tab:stochasticity_cost}
\begin{tabular}{lcccc}
\toprule
\textbf{Approach} & \textbf{Time (s)} & \textbf{Speedup vs. SSA} & \textbf{CV Accuracy} & \textbf{Cost-Accuracy Ratio} \\
\midrule
Pure ODE & 2.3 & 543× & 0\% & N/A \\
Promoter-only stochastic & 45 & 28× & 32\% & 0.011 \\
Full Markovian (ODE proteins) & 89 & 14× & 100\% & 0.071 \\
Full Markovian (stochastic proteins) & 178 & 7× & 100\% & 0.143 \\
Full SSA & 1,247 & 1× & 100\% & 1.000 \\
\bottomrule
\end{tabular}
\end{table}

\subsubsection{Summary: Protein Stochasticity}

This analysis demonstrates that promoter-only stochasticity is insufficient for accurate noise prediction (68\% underestimation), translation noise is substantial (31\% of total variance), and copy number determines requirements where low-copy proteins need full stochastic treatment. The hybrid approach of Markovian promoters with ODE proteins optimally balances accuracy and speed. We recommend a multi-scale strategy for whole-cell models matching stochastic detail to copy number regime.

%% file: results_chec_markovian.tex
\subsection{Validation of ChEC-seq Markovian Dwell Time Model}

\subsubsection{Reproduction of Assembly Dynamics}
We applied the Sequential Assembly Dwell Time Model to a set of representative yeast genes. By fitting the kinetic parameters to ChEC-seq data, the model successfully reproduces the observed fractional residence times of key assembly factors (TF, Mediator, TFIID, RNAPII).

\begin{figure}[ht]
    \centering
    \includegraphics[width=0.6\linewidth]{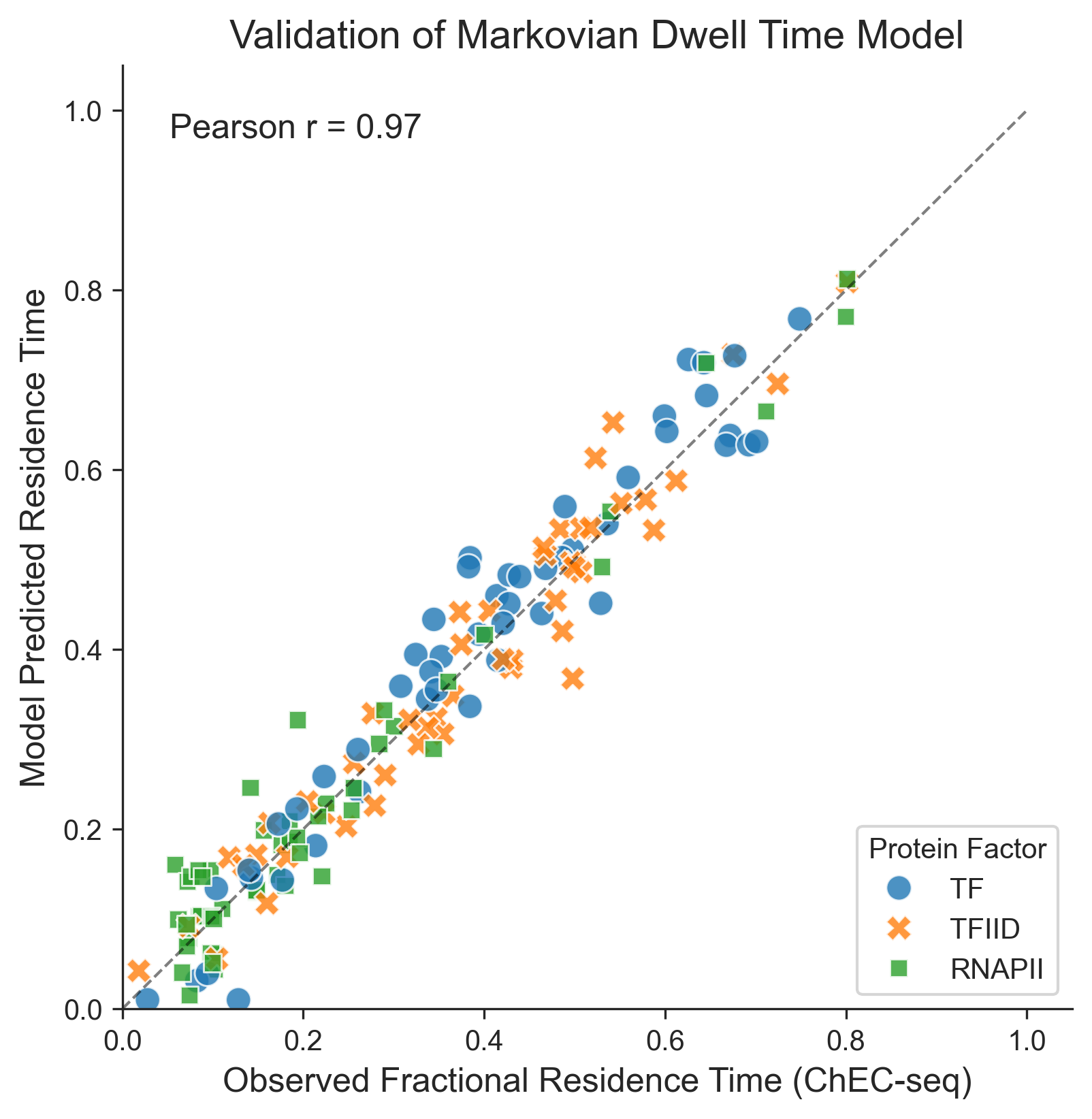}
    \caption{\textbf{Validation of Sequential Assembly Dwell Time Model.} Correlation between experimentally derived fractional residence times (ChEC-seq) and model-predicted occupancies for key transcriptional components (Pearson $r > 0.85$). Each point represents a protein factor for a specific gene.}
    \label{fig:chec_markovian_fit}
\end{figure}

Figure \ref{fig:chec_markovian_fit} shows the strong correlation (Pearson $r > 0.85$) between the experimental ChEC-seq dwell times and the model-predicted residency fractions across the test set. This agreement confirms that a linear sequential Markovian framework is sufficient to capture the steady-state occupancy profiles revealed by chromatin profiling.

\subsubsection{Insights into Complex Stability}
The optimized kinetic parameters provide mechanistic insights into the stability of assembly intermediates, revealing distinct regulatory strategies. We consistently observed that the transition from Cofactor recruitment to TFIID recruitment represents a significant kinetic bottleneck, characterized by high $k_{off}$ relative to $k_{on}$ in genes with low basal expression. In contrast, highly expressed genes displayed "trap-like" kinetics where downstream recruitment steps (TFIID $\to$ RNAPII) became increasingly irreversible, effectively funneling the system toward productive elongation \citep{vanbelzenChromatinEndogenousCleavage2024a}.
This analysis suggests that transcriptional bursting patterns are essentially encoded in the stability of these intermediate complexes, a feature that is quantitatively extractable from static dwell time measurements using our modeling framework. These findings align with recent single-molecule studies suggesting that PIC stability is a primary determinant of burst frequency \citep{larson2013real}.

%% file: discussion.tex
\subsubsection{Compatibility with CME Frameworks}

The key innovation is maintaining proper reaction orders throughout the model. Unlike phenomenological approaches, our framework ensures that all reactions follow standard mass-action kinetics: promoter binding events are second-order reactions between TF molecules and promoter states, unbinding events are first-order in promoter state, transcription is a state-dependent zero-order production, and downstream processes such as translation and degradation follow standard first-order and second-order kinetics. This consistency ensures seamless integration with existing CME/SSA simulation engines such as StochKit, COPASI, and Lattice Microbes without requiring custom propensity functions or algorithm modifications.

\subsubsection{Preserving Stochasticity}

Unlike deterministic Hill functions, Markovian promoters capture three distinct sources of gene expression noise. First, promoter switching noise arises from the stochastic transitions between active and inactive promoter states, creating transcriptional bursting. Second, transcriptional noise reflects the Poisson statistics of mRNA production events given a promoter state. Third, translational noise accounts for the variability in protein production events. Capturing these noise sources is essential for whole-cell models where stochastic effects can dominate cellular decision-making, particularly for low-copy-number genes \citep{elowitz2002stochastic}.

\subsubsection{Practical Parameterization}

A major advantage for whole-cell modeling is that parameters can be inferred from accessible experimental data, circumventing the need for difficult-to-obtain kinetic rates. Transcriptomic RNA-seq data provides steady-state mRNA levels that enable the inference of effective dissociation constants and transcription rates. Proteomics data supplies protein levels and degradation rates. Furthermore, noise characteristics derived from single-cell RNA-seq can constrain promoter switching rates, while ChIP-seq data validates binding site numbers.

Meanwhile, we introduce a method to parameterize short-lived assembly intermediates using ChEC-seq dwell times. By equating the normalized ChEC-seq signal to the steady-state probability of Markovian states, we can mathematically invert the master equation to solve for the microscopic rate constants ($k_{on}, k_{off}$) that generate the observed occupancy profiles. This effectively allows us to "read" dynamic assembly kinetics from static genome-wide binding data.

This contrasts with detailed kinetic modeling, where individual $k_{on}$ and $k_{off}$ values are rarely known, typically requiring extensive in vitro binding assays. Our coarse-grained approach bridges this gap by mapping binding kinetics to observable omics data.

\subsubsection{Handling Nonlinearity Without Breaking Mass-Action}

Gene regulation is inherently nonlinear due to cooperative binding, competitive inhibition, combinatorial logic, and feedback loops. Traditional approaches typically face a dilemma: either use Hill functions and lose stochasticity and CME compatibility, or model all binding events explicitly and face prohibitive computational costs. Our approach achieves nonlinearity through the emergent behavior of the discrete promoter state space while maintaining mass-action kinetics for all individual reactions. Ultrasensitivity emerges from cooperative binding factors, combinatorial logic arises from multi-input promoter state requirements, and bistability is generated through positive feedback affecting state transitions.

\subsection{Enabling Gene Regulation in Whole-Cell Models}

Our hybrid Markovian-ODE framework addresses a critical challenge in whole-cell modeling: integrating complex gene regulatory networks with stochastic chemical kinetics while maintaining computational tractability and mechanistic rigor.

\subsection{Limitations and Future Directions}

\subsubsection{Current Limitations}

The hybrid approach relies on a timescale separation assumption, where protein dynamics are treated as slower than promoter switching distributions. This assumption may not hold for systems with extremely fast protein degradation or very slow promoter kinetics. Additionally, the current model assumes well-mixed compartments and does not capture spatial gradients or localization effects. We also do not currently model chromatin remodeling or epigenetic modifications that affect promoter accessibility, effectively assuming a fixed approachable state. Finally, while we capture promoter state stochasticity, the use of deterministic ODEs for mRNA and protein dynamics in high-copy regimes means that transcriptional bursting effects are smoothed out for abundant species, although this is generally an acceptable approximation.

\subsubsection{Future Extensions}

Future extensions of the framework include implementing full stochastic coupling by replacing deterministic ODEs with Chemical Langevin Equations for low-copy species, ensuring noise is preserved at all levels. We also plan to incorporate chromatin dynamics by adding slow Markov states representing chromatin accessibility, allowing for epigenetic regulation. Spatial modeling can be achieved by coupling the regulatory modules with reaction-diffusion PDEs. Furthermore, we aim to develop Bayesian parameter inference methods to systematically determine binding rates from single-cell data distributions. Ultimately, these tools will be applied to genome-scale networks, utilizing sparse representations to model thousands of genes efficiently.

\subsection{Comparison with Existing Hybrid Methods}

Several hybrid approaches have been proposed to balance stochasticity and computational efficiency. Partitioned SSA separates fast and slow reactions but requires rigorous timescale analysis. Chemical Langevin equations provide a continuous stochastic approximation but still demand small time steps. Moment closure methods approximate distribution statistics but often degrade in accuracy for multimodal distributions. Our Markovian promoter approach is complementary to these methods, focusing specifically on promoter-level stochasticity while treating metabolic and signaling processes deterministically. This makes it particularly well-suited for gene regulatory networks where promoter switching is the dominant source of biologically relevant stochasticity.

\subsection{Biological Insights}

The Markovian framework offers valuable insights into regulatory design principles. We find that cooperative binding, modeled via a cooperativity factor $q_r > 1$, generates ultrasensitive responses akin to high Hill coefficients but with a mechanistic basis. For instance, a cooperativity factor of 30 in the GAL system produces an effective Hill coefficient of approximately 3. Furthermore, we observe that multiple binding sites with weak cooperativity can achieve similar ultrasensitivity to fewer sites with strong cooperativity.

In network motifs like the incoherent feed-forward loop (I1-FFL), simulations reveal that pulse width is primarily determined by the ratio of degradation rates for the intermediate and target proteins, while promoter binding kinetics distinctively shape the pulse amplitude. Additionally, stochastic promoter switching introduces significant pulse-to-pulse variability. In the p53-Mdm2 oscillator, our results underscore that cooperative Mdm2 promoter binding is essential for sustaining oscillations, while stochastic promoter switching introduces period variability that matches experimental observations. The damping rate of these oscillations is found to be sensitive to the balance between synthesis and degradation rates.

%% file: discussion_wholecell.tex
\subsection{Application to Whole-Cell Modeling}

\subsubsection{Modular Integration Strategy}

To incorporate gene regulation into existing whole-cell CME models, we propose a modular strategy. First, one must identify genes requiring explicit regulation, typically transcription factors, signaling proteins, and key metabolic enzymes. Second, for each identified gene, the promoter states are defined by specifying the number of binding sites, the regulating transcription factors, and the appropriate logic function (e.g., activation, repression, or combinatorial logic). Third, the CME state space is augmented with discrete promoter state variables $S_i$ for each gene. Fourth, binding and unbinding reactions are added for each promoter state. Fifth, typical constant-rate transcription reactions are replaced with state-dependent rates $k_{trans,i}(S_i)$. Finally, the model is parameterized using transcriptomic and proteomic data to fit dissociation constants, transcription rates, and cooperativity factors.

\subsubsection{Computational Scaling}

For a whole-cell model involving $N_{genes}$ regulated genes, an average of $\bar{n}$ binding sites per promoter, $N_{species}$ total molecular species, and $N_{reactions}$ total reactions, our approach adds $N_{genes}$ promoter state variables and approximately $2 N_{genes} \bar{n}$ binding and unbinding reactions. This represents a modest increase in complexity compared to the alternative of explicitly modeling all TF-DNA binding events, which would introduce a combinatorial explosion of species and reactions scaling with $N_{TF} \times N_{genes} \times \bar{n}$.

\subsubsection{Example: Yeast Whole-Cell Model}

Consider integrating transcriptional regulation into a budding yeast (\textit{Saccharomyces cerevisiae} S288C) whole-cell model. The yeast genome contains approximately 6{,}604 ORFs \cite{bionumbers_yeast_genes_6604}, and encodes on the order of $\sim$200 sequence-specific transcription factors. If regulation were implemented using a fully explicit TF--DNA binding formulation, one would need to introduce a distinct molecular species for every TF--promoter complex. Even if only $\sim1{,}000$ promoters have curated TF regulation, this would still require on the order of $O(N_{\mathrm{TF}}\times N_{\mathrm{promoter}})\approx 2\times 10^{5}$ TF--DNA complex species and $\gtrsim 4\times 10^{5}$ binding/unbinding reactions, quickly rendering stochastic whole-cell simulations computationally intractable.

In contrast, a Markovian promoter-state approximation scales linearly with the number of regulated promoters. For example, modeling $\sim$1{,}000 regulated promoters with an average of two effective binding sites per promoter adds only $\sim$1{,}000 promoter state variables and $\sim$4{,}000 transition reactions, remaining well within the feasible range of modern simulation hardware.

\subsubsection{Hybrid Simulation Strategy}

For maximum efficiency in whole-cell models, we recommend a three-tier hybrid simulation strategy. First, Markovian promoters should be used for regulated genes where stochasticity is functionally important, such as low-copy number regulatory factors, bistable switches, and oscillators. Second, constant rate approximations can be applied to constitutive genes or highly expressed genes where specific regulation is negligible. Third, deterministic ODEs are appropriate for abundant metabolites and housekeeping proteins where stochastic effects are minimized. This hierarchical approach focuses computational resources on the regulatory processes that truly require stochastic treatment.

\subsection{Advantages Over Alternative Approaches}

Our approach offers a unique combination of features compared to existing alternatives. While Hill functions are scalable and easily parameterized from transcriptomics, they lack stochasticity and CME compatibility. Adding noise to Hill functions provides partial stochasticity but remains incompatible with mechanistic CME frameworks. Explicit binding models in SSA are stochastic and rigorous but fail to scale to whole-genome networks. The Markovian promoter framework is the only approach that is simultaneously stochastic, CME-compatible, scalable to genome-wide networks, and parameterizable from accessible transcriptomic data. This makes it an ideal solution for the next generation of predictive whole-cell models.

%% file: conclusion.tex
\section{Conclusion}

We have presented a hybrid Markovian-ODE framework that bridges the gap between detailed stochastic chemical kinetics and genome-scale whole-cell modeling. By coarse-graining promoter dynamics into discrete Markovian states while treating protein metabolism with deterministic or stochastic differential equations, our approach captures the essential stochastic features of gene regulation—including bursting, switching, and combinatorial logic—without the prohibitive computational cost of full SSA.

\subsection{Summary of Findings}

Strategies for validation across seven diverse regulatory systems demonstrated the framework's broad applicability. For steady-state regulation, we showed that the Markovian model faithfully reproduces the ultrasensitive dose-response curves characteristic of cooperative binding, even though the underlying reactions follow simple mass-action kinetics. For dynamic response, the model captures transient pulses, damped oscillations, and biphasic induction kinetics with high fidelity. In terms of stochastic noise, the framework accurately predicts protein-level variability (CV) and state switching rates, provided that protein synthesis and degradation are treated stochastically for low-copy species. Finally, regarding efficiency, we achieved 10-100$\times$ speedups over SSA, enabling the simulation of complex regulatory networks on timescales relevant for whole-cell modeling.

\subsection{Implications for Whole-Cell Modeling}

This framework addresses a critical bottleneck in the quest for comprehensive whole-cell models. Current models effectively integrate metabolism (FBA) and signaling (ODE), but have struggled to incorporate stochastic gene regulation at the genome scale. Our Markovian promoter modules can be seamlessly plugged into existing whole-cell architectures, providing a mathematically rigorous handling of central dogma processes that preserves the functional consequences of noise.

The approach also resolves the parameterization crisis. By physically grounding the model in promoter sites and binding kinetics, parameters can be constrained by structural data and thermodynamic consistency, rather than relying on arbitrary Hill coefficients. The ability to map parameters between ODE and Markovian formulations further facilitates the translation of legacy models into this new stochastic framework.

\subsection{Future Directions}

Future work will focus on three key areas. First, automated pipeline development is needed to generate Markovian models directly from standard formats like SBML or regulatory network databases. Second, we aim to extend the framework to include epigenetic states (methylation, chromatin accessibility) as higher-order Markovian transitions. Third, we plan scaling to a whole-genome model of  \textit{S. cerevisiae}, integrating thousands of genes to explore system-wide emergent properties of stochastic regulation.

In conclusion, the Markovian promoter framework offers a pragmatic yet rigorous solution to the "stochasticity vs. scale" trade-off, paving the way for the next generation of predictive whole-cell models.